\documentclass[12pt,a4paper]{article}
\usepackage[english]{babel}
\usepackage{amssymb}
\usepackage{amsmath}
\label{eq:}\label{eq:}
\numberwithin{equation}{section}
\newcommand{\C}{\mbox{$\,${\sf I}\hspace{-1.2ex}{\bf C}}}

\begin{document}
\def\be{\begin{equation}}
\def\ee{\end{equation}}
\def\bea{\begin{eqnarray}}
\def\eea{\end{eqnarray}}
\def\mn{{\mu\nu}}
\def\ha{{1\over 2}}
\def\astf{{}^*f_{\mn}}
\def\Et{\widetilde E}
%----------------------
\def\ore{\overrightarrow{e}}
\def\orv{\overrightarrow{v}}
\def\orh{\overrightarrow{h}}
\def\ora{\overrightarrow{a}}
\def\rappD{\overrightarrow D}
\def\lappD{\overleftarrow D}
\def\orD{\rappD^R_\mu}
\def\olD{\lappD^R_\mu}
\def\simD{\buildrel\sim\over{D}}
\def\simG{\buildrel\sim\over{G}}
\def\Ft{\buildrel\sim\over{F}}
\def\p{\phi}
\def\ph{\hat\phi}
\def\Uem{U(1)_{e.m.}}
\def\gp{\bar g_\phi}
\def\Gt{\buildrel\sim\over{\Gamma}}
\def\rsimD{\overrightarrow{\simD}}
\def\lsimD{\overleftarrow{\simD}}
\def\L{{\cal L}}
\def\pa{\partial}
\def\tn{\tilde\nabla}
\def\a{\alpha}
\def\d{\delta}
\def\l{\lambda}
\def\r{\rho}
\def\s{\sigma}
\def\ga{\gamma}
\def\Ga{\Gamma}
\def\om{\omega}
\def\ka{\kappa}
\def\si{\sigma}
\def\b{\beta}
\def\w{\mbox{w}}
\def\vphi{\varphi}
\def\fnm{\phi^{(n,m)}}
\def\te{\tilde\eta}
\def\ha{\frac{1}{2}}
\def\six{\frac{1}{6}}
\def\vv{\vphi}
\def\ve{\varepsilon}
\def\lp{l_\vphi}
\def\px{\psi(x)}
\def\P{{\cal P}}
\def\H{{\cal H}}
\def\hpss{{\hps}_{s_3}}
\def\Us{U^{(s)}}
\def\Vpm{V^\pm_m}
\def\Mpm{{\cal M}^\pm_m}
\def\ua{\underline{a}}
\def\Gam{\Gamma_\mu}
\def\da{\dagger}
\def\gat{\tilde\gamma}
\def\pr{\psi_R}
\def\pl{\psi_L}
\def\pbr{\overline{\psi}_R}
\def\pbl{\overline{\psi}_L}
\def\pb{\overline{\psi}}
\def\dtt{\buildrel\approx\over\delta}
\def\gt{\tilde g}
\def\gtt{\tilde g'}
\def\gz{\tilde g_0}
\def\e{\tilde e}
\def\eh{{e\over{\hbar c}}}

\def\q{\hat q}
\def\tw{\theta_W}
\def\hph{\hat\vphi_0}
\def\WW{\hat W}
\def\ZZ{\hat Z}
%----------------new macros----------------------------
\def\role{r$\hat{\rm o}$le }
\def\ww{\rm Weyl weight }
\def\rl{\rm r${\hat {\rm o}}$le }

\def\xix{\tilde\xi(x)}
\def\xixp{\tilde\xi '(x)}
\def\xia{\tilde\xi^a(x)}
\def\xib{\tilde\xi^b(x)}
\def\xta{\tilde\xi^a}
\def\xtb{\tilde\xi^b}
\def\xicirc{\buildrel\circ\over{\tilde\xi}{\hskip-1mm}(x)}
\def\xibcirc{\buildrel\circ\over{\tilde\xi}{\hskip-1mm}^b(x)}
\def\xiacirc{\buildrel\circ\over{\tilde\xi}{\hskip-1mm}^a(x)}

\def\xxix{(x,\xix)}

\def\Ax{A_{g(x)}}
\def\Axinv{\Ax^{-1}}

\def\Axi{A\bigl(\tilde\xi(x)\bigr)}
\def\Axip{A\bigl(\tilde\xi '(x)\bigr)}
\def\Al{A\bigl(\Lambda(\xixp,\xix)\bigr)}

\newcommand{\Ag}{A_{g(x)}}
\newcommand{\tf}{{3\over 4}}

\def\Dtr{{\widetilde D^R}}
\def\pp{\pmb\psi}
\def\ppb{\bar\pp }
\def\ppbL{\ppb_L}
\def\ppbR{\ppb_R}
\def\ppx{\pp(x)}
\def\ppbx{\bar\pp(x)}

\def\gafxi{\ga_5(\tilde\xi)}
\def\gafxix{\ga_5(\xix)}
\def\gamx{\ga^\mu (\tilde\xi)}
\def\gxR{({{\ga_a\tilde\xi^a}\over R^2})}
\def\gaxi{\ga_a(\tilde\xi)}

\def\St{\widetilde S}

%===========================================end macros===========

\title{Weyl Invariant Standard Model and its Symmetry Breaking}

\author{{\it Wolfgang Drechsler}\,\footnote{e-mail: wdd@mppmu.mpg.de}  \\ 
{\normalsize  Max-Planck-Institut f\"ur Physik \footnote{ F\"oringer Ring 6, 80805 M\"unchen, Germany. \quad MPI-Publication: \bf MPP--2011--80}}}

\date{}
\maketitle
\vskip-1.0cm
\begin{abstract}
\renewcommand{\baselinestretch}{1.0}
{\small\sl
A standard model is formulated in a Weyl space, $W_4$, yielding a Weyl covariant dynamics of massless
chiral Dirac fermion fields for leptons and quarks as well as the gauge fields involved for the groups
$D(1)$\,(Weyl), $U(1)_Y{\times} SU(2)_W$\,(electroweak), $SU(3)_c$\,(colour), $SO(3,1)$\,(gravity) and
$SO(4,1)$\,(strong interaction, symmetry breaking). The dynamics is based on a gauge and Weyl invariant
Lagrangean density ${\cal L}$. Gravitation is included from the beginning as the gauge aspect of the Lorentz
group which is here extended in the hadronic sector of the model to the ten parameter $SO(4,1)$ de Sitter
group. A part of the dynamics is, as usual, a scalar isospinor field $\phi$ being a section on a bundle related
to the electroweak gauge group and to symmetry breaking. In parallel to $\phi$ on the leptonic side a
section $\tilde\xi^a$ on the hadronic side is considered as part of the dynamics, governing the symmetry breaking
$SO(4,1)\longrightarrow SO(3,1)$ and recovering gravitation in the symmetry breaking limit outside the
regions in space-time where strong interactions persist. Besides spin, isospin and helicity the Weyl weights
determine the form of the contributions of fields in ${\cal L}$. Of particular interest is the appearance of a 
current-current self-interaction of quark fields allowed by the Weyl weight changing the debate about quark masses. In a second step
the $D(1)$-Weyl symmetry is explicitly broken and a universal mass scale is established through the mass of the 
$\phi$-field appearing in the symmetry breaking Lagrangean ${\cal L}_B$. The Weyl symmetry breaking is governed
by the relation $D_\mu \Phi^2{=}0$, where $\Phi$ is the norm of $\phi$. After $D(1)$ symmetry breaking the
masses of the weak bosons and of the electron appear on the scene through the energy-momentum 
tensor of the $\phi$-field.}

\end{abstract}
\newpage
%Beginn Kapitel I ===================================================
\section{Introduction} 

\qquad     This paper is concerned with the merging of two lines of reasoning: On the one side,
a standard model theory for leptons including gravitation and formulated in a Weyl space, $W_4$,
is considered together with the subsequent breaking of the dilatation degree of freedom \cite{1}  \cite{2},
and, on the other side, the endeavour to understand various aspects of strong interactions in
analogy to gravitation, i.e. in a geometric manner by extending the Lorentz group as gauge group of gravity
to the de Sitter group,
and consider for hadrons as described in the standard model also the possibility of the presence
of a dilatation degree of freedom associated with the local de Sitter group \cite{3}\footnote {Refs. \cite{1}
and \cite{3} will be referred to as  I  and  II , respectively, in the sequel.}. Both approaches taken
together lead to a Weyl invariant formulation of the standard model without introducing masses at the
starting point, and with a subsequent explicit Weyl symmetry breaking yielding nonzero masses for
several fields involved together with the form of gravitation in terms of a Riemannian metric as we
know it from Einstein's theory.

     In I only one lepton generation was considered, and it was shown that the dilatation symmetry or
Weyl symmetry breaking of this electroweak theory for leptons reproduced the mass-giving mechanism
in a similar manner as in the standard model, and without invoking a "Higgs-phenomenon", yielding the
masses of the weak gauge bosons as well as the electron mass, and, furthermore, influenced the gravitational
coupling constant in a Brans-Dicke-like manner.

     In  II  a de Sitter space with variable curvature radius $R(x)$ was considered as fiber of an associated
bundle $\Et$ over a Riemann-Cartan space-time base, and a section on this soldered bundle $\Et$
(see Sect.\,3 below) was introduced as a vehicle for symmetry breaking mediating between domains in
space-time -- characterized in size by a length related to $R(x)$ -- possessing the full $SO(4,1)$ de Sitter gauge symmetry, and
the outside of these domains, which are characterized by the reduced subgroup symmetry, $ SO(3,1)$, 
yielding a gauge description of gravitation in the usual way in terms of Cartan's moving orthonormal
frames determining, so to speak, the geometry asymptotically  i.e.  far away from the sources of
strong interactions. The variable curvature radius of the curled up 4-dimensional local de Sitter fibers, $V'_4(x)$,
of $\Et$ -- representing an "internal" space-time of constant curvature -- was considered in  II  as a "modified Weyl"
gauge degree of freedom in the bundle geometry which, in the end and after breaking this dilatation-type degree
of freedom and fixing the length $R(x)$ to a constant $R_0$ characterizing strong interactions, would raise
the hope to see a trace of what is conventionally called {\it confinement}.

     In the present paper we combine these two approaches and formulate a standard model in a Weyl space, $W_4$,
having a hadronic sector with one generation \footnote{To compare with previous results obtained for the lepton
sector in  I  and to make the theory in a first endeavour not too complicated, we treat here only one generation
of leptons as well as quarks.} of Dirac-de Sitter left- and right-handed {\it massless} quark fields. $U_4$ torsion -- as
a large scale phenomenon, which was discussed in  II  --  is not considered in this context\,\footnote{We 
shall, however, find that de Sitter induced torsion does occur locally in isolated space-time domains as pointed out
at the end of this introduction.}. 
Frequent use is made of a coset space representation of the de Sitter group
representing the de Sitter fields as objects transforming {\it nonlinearly} under $SO(4,1)$ in a realization on the
isotropy subgroup $SO(3,1)$ of the de Sitter group -- related to the point of contact of fiber and base in the soldered
bundle $\Et$ -- with the local de Sitter spaces $V'_4(x)$ (i.e. the fibers in $\Et$) being isomorphic to
the coset space $SO(4,1)/SO(3,1)$. For spinor matter fields the corresponding covering groups $Spin(4,1)$ and
$Spin(3,1)$ are involved.

     Essentially the same situation, mathematically, arises in the leptonic sector where the electroweak gauge
group (see Eq.(1.1) below) is realized nonlinearly on the electromagnetic isotropy subgroup $U(1)_{e.m.}$ 
of $U(1)_Y\times SU(2)_W$ --- a phenomenon which is conventionally called 'spontaneous symmetry breaking' and
motivated field theoretically. (Compare in this context the discussion in App. A and B of  I .)

     It is interesting to remark that the isotropy or fixed point subgroup $U(1)_{e.m.}$ of the electroweak gauge group
characterizes the asymptotically appearing fields, in this case the electromagnetic fields, i.e. light, photons.
Analogously in the hadronic sector, with $SO(4,1)$ gauge symmetry besides $SU(3)$ colour (see Eq.(1.2) below), it
is the isotropy or fixed point subgroup $SO(3,1)$ characterizing the asymptotic fields i.e. gravitation.

     The attractive feature of the here proposed combined procedure for the leptonic ($l$) and hadronic ($h$) sectors
is the following: The standard model Lagrangean ${\cal L}  =  {\cal  L}_l + {\cal  L}_h$ defined over a Weyl
space-time base, $B {=} W_4$, involving fields given as sections of various associated and spinor bundles related
to the frame bundles (or principal bundles) $P_l(B{=}W_4,G_l)$ and $P_h(B{=}W_4,G_h)$ with the respective gauge groups
\begin{eqnarray}
G_l &=&  SO(3,1)\otimes D(1) \otimes U(1)_Y \times SU(2)_W   \\   \label{11}
G_h &=&  SO(4,1)\otimes D(1) \otimes U(1)_Y \times SU(2)_W \times SU(3)_C  \label{12}
\end{eqnarray}
can contain only field combinations allowed by the Weyl weights of the fields considered. $U(1)_Y \times SU(2)_W$ in (1.1)
is the electroweak gauge group enlarged in (1.2) by $SU(3)_c$ with c for colour for hadrons. Moreover, in the
hadronic sector the gauge group $SO(3,1)$ of gravitation has in this $W_4$-formulation been extended to
the ten parameter de Sitter group following a proposal expressed in  II  to build a bridge between strong
interactions and gravitation. In Eqs.(1.1) and (1.2)  $D(1)\simeq R^+$ is the dilatation or Weyl gauge group
isomorphic to the positive real line. 

The Weyl weights, indeed, limit the possible field couplings considerably. The total Lagrangean density $\cal L$ as well as its
parts have Weyl weight zero, i.e. are Weyl or scale invariant. The consequence of this is that {\it all fields} entering
this Weyl invariant Lagrangean {\it have to be massless fields}. In a second step the common $D(1)$ symmetry
group appearing in both the gauge groups $G_l$ and $G_h$ as a factor, has to be explicitly broken to allow for
the appearance of a universal length or mass scale in the theory. It was shown in  I  that in such an explicit
breaking of the $D(1)$ gauge symmetry there appear in the electroweak sector through the energy-momentum
tensor, $\Theta_{\mn}^{(\phi)}$, of the scalar isospinor field $\phi $ (playing the role of the Higgs field) the mass terms
for the weak boson fields $W^\pm $ and $Z^0$, and the Yukawa coupling determines the mass of the electron
as usual. Since the broken Weyl theory contains Einstein's equations of general relativity as the result of varying 
the metric in the Lagrangean formulation, it is apparent that these mass terms generated in the leptonic sector
through the $D(1)$ symmetry breaking appear on the right-hand side of the gravitational field equations and
thus act as sources for gravitational fields. We consider this result as a very physical property of
the symmetry breaking mechanism proposed in this formulation of the standard model starting from a
Weyl invariant massless scenario.

In the older literature on Weyl spaces and conformal invariance in physics and classical field theory (see, 
for example, Ref.\,\cite{3a}) it was proposed to attribute a \ww $-\ha$ to mass and allow the presence 
of mass terms in a Weyl geometric formulation of the interaction of fields. It was shown by H. Tann in his
dissertation \cite{3b} that in the presence of quantum mechanics such a treatment of mass terms leads
to contradictions and should not be considered. Only massless fields are allowed in formulating the interaction
of quantum fields in a Weyl geometric setting. Since masses do appear in nature in abundance
for the fundamental fields and particles the breaking of the Weyl symmetry is an essential part 
of the problem of establishing a realistic theoretical framework in physics when starting from a Weyl invariant theory.

      An interesting question in the present context now is, whether and how mass terms -- i.e. quark masses -- 
would appear in the hadronic sector of the theory and how the variable internal curvature radius characterizing
the local de Sitter fibers, isomorphic to $SO(4,1)/SO(3,1)$, of the mentioned bundle $\Et$ is fixed and frozen
by the subsequent $D(1)$ symmetry breaking in a unified standard model of lepton and quarks including gravitation.

     In Sect.\,2 we introduce our notation and treat the leptonic sector of the standard model formulated in a $W_4$,
reviewing the results obtained in  I  with particular focus of the \rl played by the scalar field $\phi$. In Sect.\,3 we
discuss the hadronic sector of the standard model formulated in a $W_4$, introduce the de Sitter bundle formalism
and focus on the embedding $SO(4,1)$ gauge symmetry. We investigate the \rl played by the fields $\xia$ in the
theory, determine their \ww and the way this section of $\Et$ enters the hadronic Lagrangean. Particular use is
made of the nonlinear realization [abbriviated by NL]  of the $SO(4,1)$ gauge symmetry in using the section $\xia$.
In Sect.\,4 we derive the field equations following from the hadronic Lagrangean and discuss the Weyl symmetry 
breaking. In this section the main results are presented. Finally, Sect.\,5 is devoted to some concluding remarks.

     To characterize the results obtained we like to mention: Related to the property of the fields $\xia$ as symmetry
reducing fields mediating, as mentioned above,
between regions of space-time where the full $SO(4,1)$ gauge symmetry pertains
(strong interaction domains) and the outside of these domains where only the $SO(3,1)$ gauged subsymmetry
survives (pure gravitation), we find {\it confinement of quarks} and of {\it quark currents} to isolated 
regions of space-time which are characterized geometrically by the presence of {\it confined torsion}, 
i.e. de Sitter induced torsion. This follows from the form of the Cartan connection
defined on the principal $SO(4,1)$-bundle. Secondly, the $\ga_5$-structure of the quark dynamics as obtained in Sect.\,4,
derived from the de Sitter gauge aspect, is very peculiar and has unexpected properties. 
And, thirdly, "effective" mass-like terms do appear 
automatically in the de Sitter-Dirac equations for the left- and right-handed quark fields, with these terms
being determined in size -- after $D(1)$ symmetry breaking -- by the 
constant, $x$-independent, curvature radius, $R(x)=R_0$, 
of the de Sitter fibers of the bundle $\Et$ being a bundle 
reduced in the symmetry breaking, finally,  to the bundle $E$ with fixed curvature radius $R_0$
 raised over a Riemannian space-time
base $V_4$ as defined in Sect.\,3. However, at the same time a cubic term appears in the equations for the spinor
quark fields following from a current-current self-interaction of the quark fields obtained in the hadronic
Lagrangean of Sect.\,3. This nonlinear self-interaction changes the discussion about quark masses
in a fundamental way.

\newpage
%Beginn Kapitel II==================================================

\section{The Leptonic Sector\\ of a Weyl Invariant Standard Model }

\qquad In this section we briefly review the formulation of an electroweak theory for leptons
embedded in a Weyl space or Weyl geometry \footnote{As mentioned in the introduction, we consider for
simplicity only one generation.}. The notation is the same as in Refs.\cite{1} and \cite{2} except for a
minor point regarding the matrix $\ga^5$ as explained below.

A Weyl space $W_4$ is characterized by two differential forms; a quadratic differential form, $ds^2$, as
in Riemannian geometry and a linear differential form, $\ka$ :
\be
ds^2= g_{\mu\nu}(x)\, dx^\mu\otimes dx^\nu\/;\qquad
\/\ka =\ka_\mu (x)\, dx^\mu \,.
\label{21}
\ee
A $W_4$ is equivalent to a family of Riemannian spaces $g_{\mn}(x),~\/ g'_{\mn}(x)\ldots$ and corresponding
Weyl vector fields $\ka_\r (x),\ka'_\r (x)\ldots$ related  by the transformations:
\vskip -1.0 cm
\begin{eqnarray}
g'_{\mu\nu}(x)&=&\si(x)\, g_{\mu\nu} (x)  \\ \label{22}
\kappa'_\r(x)&=&\kappa_\r(x) +\pa_\r \log\si (x)\,,\label{23}
\end{eqnarray}
where $\si (x)\,{\in}\, D(1)$, with $D(1)\,{\simeq}\, R^+$ denoting the dilatation group. The transformations (2.2)
and \eqref{23} are called {\it Weyl transformations}. Eq.\eqref{23} represents a conformal transformation or
rescaling of the metric related to the transformation \eqref{23} of the Weyl vector field $\ka_\rho \,{;} \,\rho {=}0,1,2,3$. 
Here we consider a Weyl space, or Weyl space-time, of dimension $4$ possessing Lorentzian signature
$(+,-,-,-)$ of its metric. (For the earlier history of Weyl spaces and Weyl geometry we refer to the references
quoted in \cite{2}.) 

A Weyl space $W_4$ reduces to a Riemann space $V_4$ for $\si (x){=}const$, leading to $\ka_\rho{=}const$ and ultimately
to $\ka_\rho {=}0$; a Weyl space is equivalent to a Riemann space $V_4$ if the length transfer is integrable, i.e. if
the "length curvature", $f_{\mn}$, vanishes, i.e.
\vskip -0.4 cm
\be
f_{\mn}=\pa_\mu\ka_\nu -\pa_\nu\ka_\mu =0\\.
\label{24}
\ee

A generally covariant and Weyl covariant derivative will be denoted by the symbol $D{=}dx^\mu D_\mu$. We shall 
not discriminate in the notation  (as we did in  I  ) the particular type of covariant derivative when the operator $D$ 
is applied to objects of different representation character related to various different combinations of
subgroups of the total gauge group $G_l$ of Eq.(1.1)\footnote{In  I  this was done in using one or several tildes
on $D$.}. For example, a field quantity which is unrelated to weak isospin and hypercharge, $\phi^{(n,m)}(x)$,
being covariant of degree $n$ and contravariant of degree $m$, possessing the charge $q$ and the Weyl weight
$w(\fnm)$, i.e.transforming under Weyl transformations (2.2) and (2.3) as
\be
{\fnm}'(x)=[\si(x)]^{w(\fnm)}~\fnm (x)\,,
\label{25}
\ee
has the covariant derivative
\be
 D\fnm =\nabla\fnm -\om(\fnm)\,\kappa\,\fnm +{iq\over \hbar c}\, A\,\fnm .\label{26}
\ee
with $\ka =\ka_\mu (x) dx^\mu$, $A=A_\mu (x) dx^\mu$ and $\nabla =dx^\mu \nabla_\mu$, where $\nabla_\mu$
denotes the covariant derivative with respect to the Weyl connection with connection coefficients (having
Weyl weight zero): 
\be
\Gamma_{\mn}{}^\r=\bar\Gamma_{\mn}{}^\r +W_{\mn}{}^\r =\ha g^{\r\l}
    (\pa_\mu g_{\nu\l}+\pa_\nu g_{\mu\l}-\pa_\l g_{\mn})-
    \ha(\ka_\mu\d_\nu^\r +\ka_\nu\d_\mu^\r-\ka^\r g_{\mn})\/.
\label{27}
\ee
Here $\bar\Ga_{\mn}{}^\r=\{{\r\atop\mn}\}$ are the Christoffel symbols of the Riemannian metric $g_{\mn}$
and $W_{\mu\nu}{}^\r$ denotes the Weyl addition. The connection coefficients $\Ga_{\mn}{}^\r$ referring
to a greek indexed natural base (see Eq.\eqref{21}) are symmetric in $\mu , \nu $ and are scale or Weyl invariant,
i.e. $\Ga'_{\mn}{}^\r = \Ga_{\mn}{}^\r$.

A local Lorentz basis with frame and coframe vectors $e_i$ and $\theta^i ;\, i=0,1,2,3,$ respectively, may be 
introduced by
\be
e_i=\l^\mu_i(x)\, \pa_\mu   \qquad  ;\qquad \theta^i_\mu (x) \,dx^\mu
\label{28}
\ee
with the vierbein fields $\l^i_\mu (x)$ of \ww$w(\l^i_\mu ){=}\ha$ and their inverse $\l^\mu_i (x)$ of 
\ww $w(\l^\mu_i){=}-\ha$ as seen from the relation
\be
g_{\mn} (x)=\l^i_\mu (x)\l^k_\nu (x)\eta_{ik}\quad ; \quad  \eta_{ik}=diag(1,-1,-1,-1)
\label{29}
\ee
with $g_{\mn}$ having $w(g_{\mn}){=}1$  (see Eq.(2.2)) and for the corresponding expression
for the inverse metric $g^{\mn}$ having \ww\,$w(g^{\mn}){=}-1$.

In a Cartan moving frame basis and with the connection on the Lorentz-Weyl frame bundle
\be
P_W=P_W\bigl(W_4,SO(3,1)\otimes D(1)\bigr)
\label{210}
\ee
denoted by the one-forms
\be
\om_{ik}=\bar\om_{ik} - \ha (\ka_i \theta_k - \ka_k \theta_i)
\label{211}
\ee
with $\om_{ik}{=}-\om_{ki}$ and $\ka_i{=}\l^\mu_i(x)\,\ka_\mu (x)$ and the metric part, $\bar\om_{ik}$, yielding
the Ricci rotation coefficients $\bar\Ga_{jik}$ by $\bar\om_{ik}=\theta^j\,\bar\Ga_{jik}$, the geometry in a $W_4$
is determined by Cartan's structural equations \footnote{Compare  also App. A of Ref.\,\cite{2} for the conclusions
following from Eqs.(2.12)--(2.14).}
\vspace{1.0 cm}
\be
D\theta^k \equiv d\theta^k - \om_j{}^k \wedge \theta^j - \ha\ka \wedge \theta^k =  0 \\   \label{212}
\ee
\be
d\om_{ik} + \om_{ij} \wedge \om_k{}^j  =  \Omega_{ik}  \\  \label{213}
\ee
\vspace{0.2 mm}
\be
d\ka =  f    \label{214} 
\ee
Here $\wedge$ denotes exterior multiplication of forms. Eq.\eqref{212} is Weyl and local Lorentz covariant (\ww $\ha$,
vector valued) expressing the fact that the $W_4$ has vanishing torsion. Eq.\eqref{213} is Weyl invariant and
Lorentz tensor valued with the curvature two-form, $\Omega_{ik}$, being antisymmetric in $i,k$. Finally, Eq.\eqref{214}
yields the $f$-curvature with $f\,{=}\,\ha f_{ik}\,\theta^i\wedge\theta^k\,{=}\, \ha f_{\mn}\,dx^\mu\wedge dx^\nu$
(compare \eqref{24}). 

The Weyl covariant derivative of a Dirac spinor field, $\px$, of \ww\,$w(\psi )=-{3\over 4}$ (compare \cite{2} in
this context) is given by:
\be
 D\psi (x)=dx^\mu\Bigl\{\bigl(\pa_\mu+i\Gam (x)\bigr)\psi (x)+{3\over 4}\ka_{\mu} \cdot {\bf 1}~
\psi (x)\Bigr\}\,,\label{215}
\ee
where $dx^\mu\Ga_\mu(x)$ is the spin connection one-form (see Eq.\eqref{217} below), being -- apart 
from the $D(1)$ factor -- the connection 
on the Weyl frame bundle $\bar P_W$ possessing as structural group the covering group
$\overline{SO(3,1)}=Spin(3,1)$ of the Lorentz group in $P_W$ of Eq.\eqref{210}. A Dirac spinor field, $\psi (x)$,
on a $W_4$ is defined as a section on the spinor bundle
\be 
S=S\bigl(W_4,F{=}\,\C_4\times R^+,Spin(3,1)\otimes D(1)\bigr)  \label{216}
\ee
which is a bundle associated to $\bar P_W$, possessing as fiber, $F$, a product space
$\C_4 \times R^+$ with $\C_4$ being a representation space for the Dirac spinors and with
the $R^+$ fiber related to the $D(1)$ factor of the gauge group denoting (in a particular Weyl gauge)
the local unit of length which is to be used in measuring the scalar invariant $\sqrt{\bar\psi (x)\psi(x)}$ of
\ww \,$-{3\over 4}$, with $\bar\psi (x)=\psi^\dagger (x)\ga^0$ as usual (see below)\,\footnote{In Refs.\,\cite{1}
and \cite{2} the $R^+$ part of the fiber was suppressed in the notation for the spinor bundle $S$.}. 
The field $\psi (x)$ thus possesses an invariant "length" measured relative to the section taken on the $R^+$ 
part of the local fiber determining the local Weyl gauge.
We shall assume in the following that a spin structure exists on
space-time and thus the bundle $S$ to possess global sections $\px$.

The covariant differentiation of $\px$ is defined in terms of the $Spin(3,1)$-Lie algebra valued one-form
appearing in \eqref{215}:
\be
dx^\mu\,i\Gam (x)=dx^\mu\,\l^j_\mu (x){i\over 2}\Gamma_{jik}(x) S^{ik}~;\qquad S^{ik}=
{i\over 4}[\gamma^i,\gamma^k]\/ ,
\label{217}
\ee
where $\Ga_{jik}$ are the rotation coefficients of a $W_4$ and the $\ga^i;~ i=0,1,2,3,$ are the constant $4{\times}4$
Dirac $\ga $-matrices obeying
\be
\{\ga^i,\ga^k\}=\ga^i\ga^k+\ga^k\ga^i=2\eta^{ik}\cdot {\bf 1} ,  \label{218}
\ee
with ${\bf 1}{=}{\bf 1}_4$, and the vierbein fields as well as the $\eta_{ik}$ as given in Eq.\eqref{29}. 
Moreover, the $S^{ik}$ in Eq.(2.17)
represent, as usual, the generators of the Lorentz group in the Dirac spinor representation.

In the Lagrangean presented in Eq.\eqref{229} below there will appear the $x$-dependent $\ga$-matrices, $\ga^\mu (x)$
of \ww\, $w(\ga^\mu)=-\ha $:
\be
\ga^\mu=\ga^\mu(x) =\l^\mu_i (x)\ga^i\quad\mbox{obeying}\quad
\{\ga^\mu(x),\ga^\nu(x)\}=2 g^{\mn}(x)\cdot {\bf 1}\/.
\label{219}
\ee

In the standard model the left-handed Dirac spinor matter fields (labeled with a subscript $L$) are represented as 
{\it isospinors} (I=$\ha$) with weak hypercharge Y=$-\ha$ (for leptons), forming an {\it isodoublet} $(\nu_L,e_L)$
of fermion fields $\psi_L$, and as {\it isoscalars} (I=0) with hypercharge Y=$-1$ for the right-handed Dirac fields
(labeled with a subscript R) being an {\it isosinglet} leptonic matter field $\psi_R$.

Calling the left-handed and right-handed chiral projection operators, respectively, $P_L=\ha (1-i\ga_5)$ and
$P_R=\ha(1+i\ga_5)$ with \footnote{The matrix $\ga_5$ is here defined without the factor $i$ as was done in Eq.(2.19) of  I  since
later, in the hadronic sector, the $\ga_5$ matrix will here also be related to the de Sitter spinor fields and the
relations (2.18) will have to be extended to $\{\ga^a,\ga^b\}=2\eta^{ab}\cdot {\bf 1};~a,b=0,1,2,3,5 ,$ with
$\eta^{ab}=diag(1,-1,-1,-1,-1)$ requiring $(\ga_5)^2=-1$. (See Sect.\,3 below.)}
\be
\ga_5=\ga_0\ga_1\ga_2\ga_3\, ;\qquad \ga_5{}^\dagger=-\ga_5\, ;\qquad (\ga_5)^2= -1 \label{220}
\ee
and extending in the Dirac spinor bundle $S$ introduced above the fiber $F=\C_4\times R^+$ to a fiber
$\hat F=\C_4\times \hat\C \times R^+$ with $\hat\C$ given by $\C_2$ for I=$\ha$, i.e. for left-handed fields,
and $\hat\C$ given by $\C$ for I=0, i.e. for right-handed fields, the leptonic fermion fields will be given,
ultimately, as sections on the spinor bundle
\be
\hat S = \hat S\bigl(W_4,\hat F{=}\,\C_4\times\hat\C\times R^+,\bar G_l\bigr).  \label{221}
\ee
$\hat S$ is associated to $\bar P(W_4,\bar G_l)$ being the spin frame bundle with structural group $\bar G_l$
which is obtained from (1.1) by replacing the factor $SO(3,1)$ in the gauge group by the covering group
$Spin(3,1)$.

The chiral leptonic fermion fields of \ww $-{3\over 4}$ entering the Lagrangean below are thus:
\bea
\psi_L&=&{\nu_L\choose e_L}=
  {\ha (1-i\ga_5)\psi_\nu\choose \ha(1-i\ga_5)\psi_e}\,,\quad  Y=-\ha\,;    \nonumber      \\
 \psi_R&=&e_R =\ha (1+i\ga_5)\psi_e\, , \qquad \qquad   Y=-1,
\label{222}
\eea
with their adjoints $(\bar\psi =\psi^\dagger\ga_0)$:
\bea
\pbl&=&(\bar\nu_L,\bar e_L)=\left(\pb_\nu\ha
(1+i\ga_5),\pb_e\ha (1+i\ga_5)\right), Y=\ha\,;        \nonumber    \\
\pbr&=&\bar e_R=\pb_e\ha (1-i\ga_5);\qquad\qquad Y=1\/.
\label{223}
\eea
For the scalar field $\phi$ we take as representation character
with respect to $SU(2)_W$ an {\it isodoublet}, $I=\ha$, thus yielding 
\be
\phi = \left({\vphi_+\atop\vphi_0}\right) \quad {\rm with} \quad Y=\ha\, ; 
\quad \mbox{and}\quad
\phi^\dagger =(\vphi^*_+,\vphi^*_0)\quad \mbox{with} \quad Y=-\ha
\label{224}
\ee
possessing the Weyl weight $w(\phi){=}w(\phi^\dagger ){=}-\ha$.
Here $\vphi_0$ is a neutral complex field, and $\vphi_+$
is a complex field with positive charge, obeying $\vphi^*_+=\vphi_-$\,.
The relation between electric charge, $Q$, third component of isospin, $I_3$, and weak
hypercharge is :
\be
Q=I_3+Y\/.
\label{225}
\ee

The field $\phi(x)$, representing four degrees of freedom, may be regarded as a section on the bundle
\be
E'=E'(W_4, F{=}\C_2\times R^+, G_l)    \label{226}
\ee
associated to $P(W_4,G_l)$. The square of the modulus of the scalar field is
given by the $U(1)_Y$ and $SU(2)_W$ invariant of Weyl weight $w(\Phi^2){=}-1$:
\be
\Phi^2=\phi^\dagger \phi =\vphi^*_+\vphi_++\vphi^*_0\vphi_0=|\vphi_+|^2+
|\vphi_0|^2.
\label{227}
\ee

Calling the $U(1)_Y$ gauge fields $B_\mu ;\, \mu{=}0,1,2,3$, and calling the $SU(2)_W$ gauge fields
$A^i_\mu ;\, i{=}1,2,3, \,\mu{=}0,1,2,3$, with the generators of $SU(2)_W$ given by $\ha\tau_i\,;\, i{=}1,2,3$ [with the 
summation over the index $i$ of the Pauli matrices $\tau_i$ over the range $i{=}1,2,3$ implied in Eq.\eqref{228} below]
the covariant derivative of $\phi$ is given by (with $Y{=}\ha$) :
\be
D_\mu\phi =\pa_\mu\phi +\ha\ka_\mu\cdot {\bf 1}\,\phi +{i\over 2}\tilde g
A^i_\mu\,\tau_i\,\phi + i\tilde g'Y B_\mu\cdot {\bf 1}~\phi \/.
\label{228}
\ee
where $\tilde g$ and $\tilde g'$ are dimensionless coupling constants related to the Weinberg angle (compare
Eq.(2.77) of  I ).

We are now in a position to write down a $G_l$ gauge invariant, Hermitean, Lagrangean density, $\L_l$,
containing also a dynamics of the metric $g_\mn (x)$, i.e. containing "gravitation" [modulo Weyl transformations
(2.2) and (2.3)], and being formulated in terms of only {\it massless fields} i.e. scalar, fermion and gauge fields\,
$\phi ,\,\psi_L,\, \psi_R,\, g_\mn ,\, \ka_\mu ,\, B_\mu ,\, A^i_\mu $  possessing all a definite \ww $w(\phi ){=}-\ha ,\, 
w(\psi_L){=}w(\psi_R){=}-{3\over 4}, \,w(g_\mn ){=}1$, and with the Weyl weights of all the other gauge fields
$\ka_\mu ,\, B_\mu , \,A^i_\mu $ being zero :
\begin{eqnarray}	
\L_l&=&
K\sqrt{-g}\biggl\{\ha g^{\mn}(D_\mu\phi)^\dagger
D_\nu\phi -{1\over {12}}
R_s\,\phi^\da\phi -\beta (\phi^\da\phi)^2+\tilde\a R_s{}^2
\nonumber\\ 
& &+{i\over 2}
\left(\pbl\ga^\mu{\rappD}_\mu\psi_L
-\pbl{\lappD}_\mu
\ga^\mu\psi_L\right)+{i\over 2}\left(\pbr\ga^\mu
\rappD_\mu\psi_R-\pbr\lappD_\mu\ga^\mu\psi_R\right)
\nonumber\\
& &
+\tilde\ga\,[(\pbl\phi)\psi_R+
\pbr(\phi^\dagger\psi_L)]     
-\tilde\delta\,{1\over 4}f_{\mn}f^{\mn} -
\buildrel\approx\over\delta {1\over 4}\left(F_{\mn}^i
F^{\mn}_i + B_{\mn}B^{\mn}\right)\biggr\}\/. \nonumber
\label{229}\\
\end{eqnarray}
Here the arguments $(x)$ of the fields are suppressed; 
$\ga^\mu$ are the $x$-dependent $\ga$-matrices of Eq.(2.19) with \ww \,$-\ha$. Furthermore, the curvature scalar
$R_s$ with $w(R_s)=\/-1$ is given by (see Eq.(A31) of \cite{2} )\footnote{We call the curvature scalar here $R_s$ to
discriminate it from the curvature radius $R(x)$ of the local fiber to be introduced in Sect.\,3 below.} :
\be
R_s=\bar R-3\bar\nabla^\r \ka_\r +{3\over 2} \ka^\r\ka_\r\,,
\label{230}
\ee
where $\bar R$ is the curvature scalar of the Riemannian space-time, $V_4$, and $\bar\nabla^\r$ denotes the
purely metric covariant derivative. $F^i_\mn ; i=1,2,3$ is the $SU(2)_W$ gauge curvature tensor and $B_\mn$
is the $U(1)_Y$ gauge curvature tensor (both antisymmetric in $\mu,\nu $) ; $\sqrt{-g}$ is the root of the
determinant of the $g_\mn$ having \ww 2, and $K$, finally, is an overall constant of dimension [Energy$\cdot L^{-1}$],
with $L$ standing for length, converting the length dimension of the curly brackets in Eq.\eqref{229}, which is [$L^{-2}$], into
[Energy$\cdot L^{-3}$] in order to give to $\L_l$, later after symmetry breaking, the correct dimension of
an energy density \footnote{The factor $K$ drops out of the field equations derived from $\L_l$ in  I  and appears here only for convenience. However, in the hadronic sector, discussed in Sect.\,3 below, it is to be assured that the same conventions
are adopted for $\L_h$ to yield the total Lagrangean $\L =\L_l+\L_h$.}. 

The various terms in the curly brackets of \eqref{229} have the following meaning: The first term is the kinetic
term of the $\phi$-field; the second term corresponds to the Einstein-Hilbert action in general relativity, with the
factor $1\over {12}$ in front guaranteeing conformal invariance in the $W_4 \rightarrow V_4$ limit; the third term would correspond to a cosmological constant, it is a self-coupling term of the $\phi$-field allowed by the \ww $w(\phi){=}-\ha$
and is  multiplied in (2.29) by a constant $\beta$ of length dimension [$L^{-2}$]; the fourth term with $R_s{}^2$
multiplied by a constant $\tilde\a$ of dimension [$L^2$] was introduced in  I  to yield a nontrivial dynamics for the
Weyl vector field $\ka_\mu$ -- being a field , which, after $D(1)$ symmetry breaking, would be removed from the scene,
thus the $\tilde\a$-term could be eliminated right at the start; however we keep it because it was included in  I   -- .
Furthermore, the fifth and sixth terms are the kinetic terms for the left-handed and right-handed leptonic spinor
matter fields; the seventh term with constant $\tilde\ga$ of dimension [$L^{-1}$] is the Yukawa-like coupling of
the $\phi$-field to the fermions. And, finally, the last three terms, with constants $\tilde\delta$ and 
$\buildrel\approx\over\delta$ of dimensions [$L^2$], are the quadratic curvature invariants of the $D(1)$,
$SU(2)_W$ and $U(1)_Y$ gauge fields, respectively. We remark in concluding that the length dimension of the scalar
field is taken to be [$L^0$], and, relative to this choice, the leptonic fermion fields have length dimension [$L^{-\ha}$],
which leads, taken together, to the above-mentioned length dimensions for the constants $\tilde\ga$ and $\beta$. 

For further comments and the derivation and discussion of the field equations following from this purely leptonic
Lagrangean density for a Weyl electroweak theory and its subsequent $D(1)$ symmetry breaking as well as for the
mass generation of the gauge bosons $W^\pm$ and $Z^0$ and of the electron, see the results presented in  I  . We
here continue completing the description of the standard model in studying the hadronic sector in a similar manner
basing it, at the start, on a Weyl geometry and consider again -- in a second step -- a $D(1)$ symmetry breaking
leading, finally, to the usual description of gravitation in a Riemannian space-time, as described in  I, together 
with the results found in Sect.\,4 below concerning the question
how nonzero masses for the basic hadronic fields do appear in the theory.

The literature on the standard model is truly enormous. At the beginning of all this development stands Weinberg's
'Model of Leptons' of 1967 \cite{4}. For a more recent account of the history of the standard model up to the end
of 2003 we refer to Weinberg's talk given at CERN \cite{5} containing numerous references to the original
literature marking a long and exciting journey.

%Beginn Kapitel  III ============================================================

\section{The Hadronic Sector\\ of a Weyl Invariant Standard Model}

\qquad In the sixties of the last century several members of the physics community expressed their doubts
concerning the usefulness of quantum field theoretic methods for the description and understanding of strong
interactions. Despite the great successes of quantum field theory in QED the situation in strong interaction
physics looked less promising. There was the Regge-pole model with its complex angular momentum and
interpolating fields as well as the use of dispersion relations, which brought a limited amount of success and 
progress. The main highlight and focus, however, was {\it symmetry}, motivated by group theory and
representation theory, in particular $SU(3)$ and Gell-Mann's "Eightfold Way" \footnote{See, for example Ref.\,\cite{6}},
which, in the end, led to quarks and the colour-$SU(3)$ description which is now part of the standard model.

Some physicists followed in the seventies and early eighties an altogether different route trying to link
strong interactions to gravity. They studied the possibility of regarding the Poincar$\acute{\rm e}$ group
as a gauge group with the hope to arrive at a dynamical theory involving a spectrum of states of various
masses and spins in the presence of gravitational fields and, possibly, new geometric additions to Einstein's theory
as, for example, torsion (compare the literature quoted in Ref.\,\cite{7}). In this general context it was
suggested in 1975  to use differential geometric means and methods and to base a geometric description
for extended elementary objects in particle physics
on a higher dimensional space, in fact, a {\it fiber space} or {\it  fiber bundle}
raised over space-time as base \cite{8}. In particular, it was proposed to use a bundle with the ten parameter
de Sitter group, $G{=}SO(4,1){=}O(4,1)^{++}$, as gauge or structural group, having the gauge group of gravitation, 
$H{=}SO(3,1)$, as a subgroup, and to consider as fiber of such a bundle a coset space $G/H$ isomorphic to a
Riemann space, $V'_4$, of constant curvature. The curvature radius $R$ was considered in \cite{8} to be a constant
(i.e. to be independent of $x{\in} B$) characterizing hadron physics, i.e. being of the order of $R{\sim}10^{-13}$ cm
with $R$ representing a length parameter typical for extended elementary objects like protons or neutrons.

An essential point in this approach is the use of a so-called {\it soldered bundle} (here of eight dimensions)
possessing a first order contact between the local fiber, $V'_4$, at $x{\in}B$ and the space-time base $B$ being
a Riemannian space $V_4$. The soldering property\footnote{Compare Ref.\,\cite{9} and the literature quoted there.}
leads to the identification of the local tangent space to space-time and the local tangent space to the fiber -- being
both 4-dimensional Minkowski spaces -- by an isomorphism. The particular bundle proposed in Ref.\,\cite{8} was the 
de Sitter bundle :
\be
E=E\bigl(B{=}V_4,F{=}G/H,G{=}SO(4,1)\bigr)
\label{31}
\ee
which is associated to the de Sitter frame bundle $P{=}P(V_4,G{=}SO(4,1))$ over space-time $V_4$.

In contracting the structural group in $E$ and $P$ by an In\"on\"u-Wigner contraction with respect to the
stability subgroup $H$, and considering the limit $R\longrightarrow\infty$, the de Sitter group $SO(4,1)$ 
contracts to the Poincar$\acute{\rm e}$ group, $ISO(3,1)$, and $E$ goes over to the affine tangent bundle over
space-time $V_4$ associated to the Poincar$\acute{\rm e}$ frame bundle $P(V_4,ISO(3,1))$ over $V_4$ \cite{10}.

Some years later, in the paper  II  \footnote{See also Ref.\,\cite{11}}, the curvature radius $R$ characterizing the local de Sitter fibers, was considered to be a variable, i.e.\,$x$-dependent, quantity, $R{=}R(x)$, referred to in  II  as a "modified Weyl" degree of freedom. It was named in this way, since it applied to the variability of a quantity associated with
the {\it fiber} of a bundle over space-time, which was unrelated in  II  to the metric of the underlying space-time base.
Now we intend to extend the Weyl transformations (2.2) and \eqref{23} considered above also to the
hadronic sector. Now the curvature radius of the local de Sitter fiber does, indeed, transform under the Weyl
transformations introduced in Sect.\,2 above, and we have to determine a \ww \footnote{The \ww which was introduced
in the "modified Weyl" theory in Eq.(1.3) of  II  has to be changed now.} for the quantity $R(x)$.
Before we do that we enlarge our above de Sitter framework with the associated bundle $E$ to a bundle $\Et$
over a base $W_4$ having the structural group $SO(4,1)\otimes D(1)$ (calling the bundle again $\Et$ as in  II  ):
\be
\Et = \Et \bigl(B{=}W_4,\widetilde F{=}G/H\times R^+,SO(4,1)\otimes D(1)\bigr)
\label{32}
\ee
which is a bundle associated to the principal bundle $\widetilde P_W(W_4,SO(4,1)\otimes D(1))$ called 
the de Sitter-Weyl frame bundle.
We stress that we shall here not consider the possibility of the global presence of Cartan $U_4$ torsion in the base,
as was considered essential in  II \footnote{In comparing formulae appearing here with those derived in II the 
torsion tensor $K^R_{ijk}$ there has to be put to zero.}. In Eq.\eqref{32} we regard the local fiber, $\widetilde F_x$
over $x{\in} B$, to be the direct product of a local de Sitter space-time, $V'_4\simeq G/H$, as before in $E$, and the
positive real line, $R^+$, determining the local curvature radius $R(x){\in}R^+$ in considering sections on $\Et$. 
We parametrize the de Sitter space $V'_4$ in terms of coordinates \,$\xta$\,; $a{=}0,1,2,3,5$ of an embedding Lorentzian 
space $R_{4,1}$ with metric $\eta_{ab}{=}diag(1,-1,-1,-,1-1)$, with the space $V'_4$ given in $R_{4,1}$ as a
hypersurface, i.e. a one-shell hyperboloid $\xta\xtb\eta_{ab}{=}-R^2$, with the summation convention of
summing over repeated upper and lower indices $a,b$ over the range $0,1,2,3,5$ being understood\,\footnote
{antipodal points $\xta$ and $-\xta$ are identified on the hyperboloid.}. Thus a section on $\Et$ will be denoted
by $(\xia,R(x))$ obeying
\be
\xia\,\xib\,\eta_{ab}=-R^2(x) ,
\label{33}
\ee
with the $\xia$ transforming under $SO(4,1)$ gauge transformations, i.e.\,changes of section on $\Et$
 (disregarding the $D(1)$ transformations at this point), as
\be
{\tilde\xi^a}{}'(x)=[A_{g(x)}]^a{}_b\, \xib ,
\label{34}
\ee
where $A_{g(x)}$ denotes an $x$-dependent element of $SO(4,1)$ in its basic $5\times 5$ matrix representation.

The point of contact between base space and fiber in the soldered bundle $\Et$ will be denoted by $\xicirc$ having
coordinates
\bea
\xibcirc &=&\bigl(\xibcirc = 0 \,\, {\rm for}\,\,  b{=}i{=}0,1,2,3\, ; 
\,\, \xibcirc = -R(x) \,\, {\rm for}\,\, b{=}5 \bigr)  \nonumber \\
            &=& \bigl(0,0,0,0,-R(x)\bigr) .
\label{35}
\eea
The de Sitter boost transferring the point $\xicirc$ on $\Et$ to the point $\xia$ on $\Et$ is denoted by $\Axi$
obeying
\be
\xia = \bigl [\Axi\bigr ]^a{}_b\,\xibcirc .
\label{36}
\ee
An explicit matrix representation of $\Axi$ is given in Eq.(2.20) of II (see also App. A below).

The Lorentz group $H$ in the discussion above concerning the bundles $E$ and $\Et$ is taken to be the isopropy
subgroup of $SO(4,1)$ belonging to the point $\xicirc$ in $G/H{\simeq}V'_4(x)$. Each de Sitter transformation,
$A_{g(x)}$, may now be broken down into two boosts and a Lorentz transformation in $\xicirc$, which may be 
expressed as :
\be
A_{g(x)}\,\Axi = \Axip \,\Al
\label{37}
\ee
with $\tilde\xi^a{}'(x)$ and $\xia$ related by Eq.\eqref{34}. The transformations $\Al$ represent the
{\it nonlinear realization} of the de Sitter transformations on the stability or isotropy subgroup $H$ of
the point $\xicirc$ in $\Et$, which are the so-called $SO(4,1)$ "Wigner rotations", being Lorentz transformations in 
this case.

We now return to the question of the \ww for $R(x)$, or more specifically, for the section $(\xia ,R(x))$ of $\Et$. 
The intrinsic geometry of a $(4,1)$-de Sitter space with curvature radius $R$ was investigated in stereographic
projection [sp] coordinates $\tilde x^i , i{=}0,1,2,3,$ in Ref.\,\cite{12} and shown to be determined by Cartan's
structural equations :
\be
\tilde d\tilde \om^i+\tilde\om_k{}^i\wedge\tilde\om^k=0\quad{\rm and}\quad\tilde d\om_{ij}+\tilde\om_{ik}\wedge\tilde\om_j{}^k=
\tilde\Omega_{ij}={1\over{R^2}}\,\tilde\om_i\wedge\tilde\om_j
\label{38}
\ee
with the metric on the de Sitter space being expressed in terms of [sp] coordinates by $\tilde g_{\mn}(\tilde x)$
(see App.\,A of \cite{12}). This metric leads to a curvature scalar $\tilde R_s$ for a de Sitter space which, together
with the curvature equation in \eqref{38} implies the relation
\be
\tilde R_s={32\over R^2}=const .
\label{39}
\ee
Because of soldering the internal  metric $\tilde g_{\mn}(\tilde x)$ in the de Sitter fiber of $\Et$ has to be scaled 
in the same way as the metric in the base (see Eqs.(2.2) and (2.3) of Sect.\,2 above), implying that $\tilde R_s$
has \ww $w(\tilde R_s){=}-1$ as has the curvature scalar in Eq.\eqref{229}. The Eq.\eqref{39} then shows
that the curvature radius $R(x)$ of the fiber of $\Et$ is transforming with \ww $w(R){=}\ha$, i.e. 
\be
R'(x)=[\si(x)]^{\ha}\,R(x) .
\label{310}
\ee
The same is true for $\xia$ transforming under Weyl transformations (2.2) and (2.3) with \ww $w(\xi){=}\ha$.

This information on the \ww for the sections on $\Et$ is crucial in order to determine how these sections
$(\xia, R(x))$ --- which eventually will determine the symmetry breaking $SO(4,1)\longrightarrow SO(3,1)$
and the \role  played by the asymptotically surviving gravitational fields --- will enter the hadronic Lagrangean
and thereby enter the dynamics of strong interactions. The section $\xia$, with four degrees of freedom, is here 
completely analogous to the scalar field $\phi(x)$ in the leptonic sector studied in Sect.2. We shall come back to 
this analogy in Sect.4 below.

Next we like to discuss the connection defined on $\widetilde P_W(W_4,SO(4,1)\otimes D(1))$ determining the covariant
differentiation on the associated bundles $\Et$ and $\tilde S$ below characterizing the strong interactions
together with gravitation geometrically in the hadronic sector of the standard model. This means that we are
going to treat the quark fields, on the one hand, as left-handed and right-handed {\it massless} Dirac
fermion fields with the conventional weak isospin and hypercharge assignments as well as the $SU(3)_c$ gauge
interaction; on the other hand, however, as four component de Sitter spinor fields transforming under 
$\overline {SO(4,1)}{=}Spin(4,1)$ at small ("confined") distances and under $Spin(3,1)$ asymptotically, as
we shall see. The $D(1)$ factor in the bundle $\widetilde P_W$ introduced above -- which will be explicitly
broken later as mentioned -- is treated as in Sect.\,2 (compare Eqs.\eqref{210}{--}\eqref{214}) with the tilde on
$\widetilde P_W$ understood as a reminder that the de Sitter group is involved here as compared to Eq.\eqref{210} 
[and that the $D(1)$ group is present in distinction to Eq.\eqref{31}].  However, we focus the attention primarily
on the de Sitter degrees of freedom in the following discussion.

We now introduce a so-called Cartan connection \footnote{Compare Ref.\,\cite{9} for a more mathematical
treatment.} on the soldered bundle $\Et$ possessing a curled up fiber $V'_4{\simeq}SO(4,1)/SO(3,1)$ of variable
radius of curvature $R(x)$. The connection on $\widetilde P_W$ to which $\Et$ is associated (disregarding for
the moment, as mentioned, the dilatation group) is a $SO(4,1)$-Lie algebra-valued one-form, $\om^R$, which,
in a particular gauge, gives rise to a $5\times 5$ matrix of one-forms in the base of $\widetilde P_W$, i.e.\,\footnote
{See Ref.\,\cite{9}. The inhomogeneous transformation rule under $SO(4,1)$ gauge transformation, $\Ax$, of
the matrix (3.11) is given by: $\om^R(x)'{=}\Ax\om^R(x)\Axinv - \Ax d\Axinv$.}  
\be
[\om^R(x)]^a{}_b =
\begin{pmatrix}
 [\om^R(x)]^i{}_j & {\theta^R(x)}^i \cr
{\theta^R (x)}_j  &  0                     \cr
\end{pmatrix}
\label{311}
\ee
Here $a{=}i,5$ (row index) and $b{=}j,5$ (column index) with $i,j{=}0,1,2,3$. Lowering the index $a$ with
the de Sitter metric $\eta_{ab}$ one obtains the forms $\om^R_{ab}$ obeying $\om^R_{ab}{=}-\om^R_{ba}$ 
\footnote{The label $R$ on the connection forms is a reminder that a de Sitter space of curvature radius $R$
is involved.}.

Calling now the de Sitter Lie algebra {\bf g}, the Lorentz subalgebra {\bf g$\prime$} and the de Sitter boost
generators {\bf p}, the form \eqref{311} of the connection corresponds to the decomposition of the Lie algebra
{\bf g} according to {\bf g}={\bf g$\prime$}+{\bf p} with the {\bf p}-valued forms denoted by $\theta^R(x){}^i$
and $\theta^R(x){}_j$, respectively,  representing the soldering forms of the connection.

The geometry in $\widetilde P_W(W_4,SO(4,1)\otimes D(1))$ is characterized by Cartan's structural equations 
reading
\be
d\,\om^R_{ab} + \om^R_{ac}\wedge\om^R_b{}^c = \Omega^R_{ab}
\label{312}
\ee
The \role of the $SO(3,1)\otimes D(1)$ subsymmetry was studied in Eqs.\eqref{210} -- \eqref{214}, and the
relations to the Eqs.\eqref{312} have now to be established. To this end we go over to the nonlinearly [NL]
transforming form of the connection \eqref{311} in using the boost transformations $\Axi$\,\footnote{For details
see  II, p. 630 - 632.}. The result is the following NL form for the de Sitter connection :
\be
[W^R\xxix]^a{}_b =
\begin{pmatrix}
[W^R\xxix]^i{}_j & [\theta^R\xxix]^i  \cr
[\theta^R\xxix]_j & 0                             \cr
\end{pmatrix} .
\label{313}
\ee
The arguments $\xxix$ are here meant to indicate that the entries of the matrix \eqref{313} transform under
$\Al$, i.e. nonlinearly under $SO(4,1)$, with $[W^R]^i{}_j$ as a Lorentz connection and $[\theta^R]_j$ and 
$[\theta^R]^i$ as co- or contravariant Lorentz vectors.  

The transformed soldering forms in \eqref{313} are given by
\be
[\theta^R\xxix]^i\equiv[W^R\xxix]^i{}_5 = {1\over {R(x)}}
[A^{-1}\xix]^i{}_a\,\Dtr\xia ,
\label{314}
\ee
where $\Dtr\xia$ is the $SO(4,1)\otimes D(1)$ covariant derivative of the section $\xia$ on $\Et$ defined by
\be
\Dtr\xia = d\,\xia+[\om^R(x)]_b{}^a\,\xib - w(\tilde\xi)\,\ka(x)\,\xia
\label{315}
\ee
with the \ww $w(\tilde\xi){=}\ha$ as derived above.

The form \eqref{314} for $[\theta\xxix]^i$ exemplifies explicitly the fact known from differential geometry\,\footnote{
Compare the theorems on ps. 57 and 88 of Ref.\,\cite{13}.} that the connection on $\widetilde P_W$ reduces
from a {\bf g}-valued to a {\bf g$\prime$}-valued, i.e. Lorentz-valued, form whenever the section $(\xia,R(x))$
on $\Et$ is covariant constant (or {\it parallel}), i.e. obeys
\be
\Dtr\xia = 0 .
\label{316}
\ee
This equation characterizes the region in the space-time base of the bundles concerned where the original
de Sitter gauge symmetry reduces to the gauged Lorentz subsymmetry (representing pure gravitation physically)
and discriminates it from those regions in the space-time base, i.e. for $\Dtr\xia\not=0$, where the $SO(4,1)$
gauge symmetry does {\it not} reduce but is nonlinearly realized on the stability subgroup $SO(3,1)$ of
$SO(4,1)$.

In order to make the conclusions of Eq.\eqref{316} more transparent and explicit we view this equation
in its nonlinear transforming form :
\be
\buildrel {NL}\over\Dtr\,\xiacirc\equiv d\xiacirc + [W^R\xxix]_b{}^a\,\xibcirc- \ha\ka(x)\xiacirc=0 .
\label{317}
\ee
Here $\buildrel {NL}\over\Dtr$ denotes the de Sitter and Weyl covariant derivative with respect to the
nonlinearly transforming de Sitter connection $W^R\xxix$. Due to Eq.\eqref{35} the Eq.\eqref{317} is
equivalent to
\be
[W^R\xxix]^i{}_5 \equiv [\theta^R\xxix]^i =0 \quad {\rm for}\,a{=}i{=}0,1,2,3,  \\
\label{318} 
\ee
and to
\be 
d R(x)-\ha\ka(x)R(x)\equiv \widetilde D\,R(x) =0  \quad {\rm for}\, a{=}5 .
\label{319}
\ee
Eq.\eqref{318} restates the already mentioned fact that the de Sitter connection reduces from a {\bf g}-valued
to a {\bf g$\prime$}-valued form if $\Dtr\xia{=}0$, and Eq.\eqref{319} states that $R(x)$ is Weyl 
covariant constant implying that $\ka_\mu(x){=}2\pa_\mu{\rm log} R(x)$, i.e. that $f_{\mn}{=}0$ and the transfer of 
the length $R(x)$ is integrable. The Weyl degree of freedom of $R(x)$ is 
thus "pure gauge" and may, in fact, be gauged away
using in Eq.(2.3) a gauge factor $\si(x)^{\ha}{=}const/R(x)$ yielding $\ka'_\mu{=}0$ and $R'(x){=}R_0{=}const$ 
whenever or wherever \eqref{317} is true. Hence the de Sitter and Weyl gauge symmetries are both broken in
regions of the space-time base of the bundles where $\Dtr\xia{=}0$.

In view of the results obtained in  II  and in the absence of global $U_4$ Cartan torsion in the
space-time base, we are entitled to identify the Lorentz i.e. {\bf g$\prime$}-valued part in \eqref{313}, after lowering 
the index $i$, with the forms $\om_{ik}=\om_{ik}(x)$ in Eq.\eqref{211} of Sect.\,2. This means :
\be
[W^R\xxix]_{ik} = \om_{ik}(x)  ,
\label{320}
\ee
where $\om_{ik}(x)$ is composed of a purely metric part, $\bar\om_{ik}(x)$, and a Weyl addition which vanishes
whenever Eq.\eqref{316} is true. The actually remaining de Sitter contributions thus only arise from the
soldering forms $[\theta^R\xxix]^i$ in Eq.\eqref{313} leading to the presence of de Sitter-induced torsion -- i.e.
$(i,5)$- and $(5,j)$-components of de Sitter curvature -- 
in regions where $\Dtr\xia\not=0$ (see below).

We now, finally,  introduce the de Sitter spinor bundle $S'$ associated to 
$\bar{\widetilde P}_W(W_4,Spin(4,1)\otimes D(1))$ with struclural group $Spin(4,1)\otimes D(1)$, i.e. (compare Eq.
\eqref{216})
\be
S' = S'\bigl(W_4,F{=}\C_4\times R^+,Spin(4,1)\otimes D(1)\bigl)
\label{321}
\ee
with the $SO(4,1)$ spin connection, denoted by $i\Gam^R(x)$, being in analogy to Eq.\eqref{217} defined by
\be
dx^\mu i\Gam^R(x) = dx^\mu \l^i_\mu(x)\, {i\over 2}\Gamma^R_{iab}(x)\,S^{ab} ,\qquad S^{ab}={i\over 4}[\ga^a,\ga^b] ,
\label{322}
\ee
where $\ga^a{=}(\ga^i,i{=}0,1,2,3; \,\ga^5{=}\ga^0\ga^1\ga^2\ga^3)$ is a set of five anticommuting 
constant Dirac $\ga$-matrices obeying
\be
\{\ga^a ,\ga^b\} = 2\, \eta^{ab}\,{\bf 1} , \qquad  {\ga^a}^\dagger = \ga^0\ga^a\ga^0 ,
\label{323}
\ee
with the metric $\eta^{ab}$ in the embedding space $R_{4,1}$ as introduced above (compare also Eqs.\eqref{218}). 
The $\Gamma^R_{iab}$ in \eqref{322} are the de Sitter rotation coefficients with $\theta^i \Gamma^R_{iab}{=}
\om^R_{ab}$.

The $Spin(4,1)\otimes D(1)$ covariant derivative of a de Sitter spinor wave function $\pmb\psi$ of four component 
Dirac type \footnote{There are no two component Weyl spinors transforming under the de Sitter group.}
, being a section on $S'$ and possessing \ww $w(\pmb\psi){=}-\tf$ as in Sect.\,2 above, is given by 
\be
D^R\ppx = dx^\mu\bigl\{ \bigl(\pa_\mu + i\Gamma^R_\mu(x) \bigr)\,\ppx + \tf \ka_\mu(x)\cdot {\bf 1}\,\, \ppx\bigr\} .
\label{324}
\ee
Under changes of section on $S'$ , i.e. under de Sitter and $D(1)$ gauge transformations, the 
spinor field $\ppx$ transforms as
\be
{\ppx} ' = [\si(x)]^{-\tf} \,S(\Ag)\,\ppx ,
\label{325}
\ee
and correspondingly for $\ppbx{=}\pp^\dagger\ga^0$ :
\be
{\ppbx}' =[\si(x)]^{-\tf} \,\ppbx\, S^{-1}(\Ag ) ,
\label{326}
\ee 
where $S(\Ag){\in} Spin(4,1)$ and $S^{-1}(\Ag){=}\ga^0 S^\dagger (\Ag)\ga^0$ obeying
\be
S(\Ag)\, \ga^a \,S^{-1}(\Ag) = [A^{-1}_{g(x)}]^a{}_b\,\ga^b = [\Ag]_b{}^a\,\ga^b
\label{327}
\ee
expressing the homomorphism $Spin(4,1)\longrightarrow SO(4,1)$. 

It is easy to go over now to nonlinearly transforming spinor fields  with
the help of a spinor boost transformation (we use the same notation as in Eq.\eqref{313} above)
\be
\pp\xxix =  S^{-1}\bigl(\Axi\bigr)\,\ppx
\label{328}
\ee
yielding a four component spinor field transforming under changes of section on $S'$ and, correspondingly, on
$\Et$ as
\be
\pp '(x, \tilde\xi{}'(x)) = [\si(x)]^{-\tf}\,\, S\Bigl(\Al\Bigr)\,\pp\xxix ,
\label{329}
\ee
with
\be
\xta{}'(x) = [\si(x)]^{\ha}\,\,[\Ag]^a{}_b\,\xib .
\label{330}
\ee
We end this discussion by introducing, with the help of the spinor boost, a set of nonlinearly transforming
$\ga^a$-matrices under $Spin(4,1)$ by (compare \cite{11})
\be
\ga^a(\xix) = S\bigl(\Axi\bigr)\, \ga^a\, S^{-1}\bigl(\Axi\bigr) = [A^{-1}(\xix)]^a{}_b\, \ga^b  ,
\label{331}
\ee
with
\be
\ga^a(\xix) = \Bigl(\ga^k(\xix) ;\,\, \ga^5(\xix){=}{\ga_b\,\xib\over R(x)} \Bigr) .
\label{332}
\ee
As is easily shown using Eqs.\eqref{37} and \eqref{327} they, transform like
\be
S(\Ag) \,\ga^k(\xix) \, S^{-1}(\Ag) = [\Lambda^{-1}(\xixp,\xix)]^k{}_i \,\,\ga^i(\xixp)\, ;\quad{\rm for}\, a{=}k\,,
\label{333}
\ee
and \footnote{One can rewrite Eqs.\eqref{333} and \eqref{334} in one formula 
as : \\
$S(A_{g(x)})\,\ga^a(\xix)\,S^{-1}(A_{g(x)})=[A\bigl(\Lambda^{-1}(\tilde\xi '(x),\tilde\xi (x))\bigr)]
^a{}_b\,\ga^ b(\xixp)$.} 
\be
S(\Ag)\,{\ga_b\,\xib\over R(x)}\,S^{-1}(\Ag) = {\ga_b\,\tilde\xi^b{}'(x)\over R(x)}\,;\quad{\rm for}\, a{=}5\,.
\label{334}
\ee

One could regard $\pp(x)$ as a spinor field defined over the bundle $\Et$ being associated with the point
$\xxix$ there. Changing the section on $\Et$ by a gauge transformation $A_{g(x)}$ with $\xix\longrightarrow\xixp$
according to Eq.\eqref{34} -- or better \eqref{330} -- 
corresponds to the change with $S(A_{g(x)})$ for $\pp(x)$ on $S'$ in \eqref{321}
according to Eq.\eqref{325}. Going over to the nonlinearly transforming spinor fields \eqref{328} transforming as
given in Eqs.\eqref{329} and \eqref{330}, one sees that the gauge transformation on $\Et$ is still in the background.
Here $\pp\xxix$ -- despite its notation -- could be regarded as a field defined over the base of $\Et$ in
identifying the zero section $\xicirc$ on $\Et$ with the space-time base. It is essential to remark that the
transition \eqref{328} from the linear to the nonlinear transforming de Sitter spinor field is a {\it gauge transformation}.
The $\xix$ in the argument of the $\pp$-field in \eqref{328} is meant merely as a reminder or label that this is the spinor
field represented in the {\it nonlinear gauge}, i.e. transforming nonlinearly under $Spin(4,1)$. Correspondingly, 
the $\ga^a(\xix)$ denote the set of de Sitter $\ga^a$-matrices {\it in the nonlinear gauge}, i.e. transforming nonlinearly
under $Spin(4,1)$. This property is of importance later in Sect.\,4 in considering variations of the fields entering the
Lagrangean with the aim at deriving field equations for the fields $\xia$. The gauge choice for the $\ga^a$-matrices
or the spinor fields is in this respect clearly not of dynamical relevance. 
  
\vskip0.5cm

Before we finally introduce the de Sitter quark fields in this formulation of the standard model based on a $W_4$, we
drop some remarks concerning the curvature tensor associated with the two-forms $\Omega^R_{ab}$ of
Eq.\eqref{312}. Considering the NL form of this equation and in view of Eqs.\eqref{313} and \eqref{318} 
one finds that, despite of the fact that we excluded a general Cartan $U_4$ torsion in the space-time base, there does
appear, as mentioned above, a so called de Sitter-induced torsion of the Cartan connection on $\widetilde P_W$
locally in regions where $\Dtr\xia\not=0$. 
Using also the identification \eqref{320} this torsion is given in NL form by the Lorentz vector valued two-form
\bea
[\Theta^R\xxix]^i &=& d\,[\theta^R\xxix]^i + [\om(x)]_k{}^i \wedge [\theta^R\xxix]^k  \nonumber \\
                               &=& D[\theta^R\xxix]^i
\label{335}
\eea 
expressible in terms of the de Sitter connection coefficients $\Gamma^R_{ji5}\xxix$ which are Lorentz 
tensor-valued quantities here. These coefficients of the soldering forms also make a quadratic contribution
to the de Sitter curvature tensor $R^R_{ijkl}\xxix$ in NL form which reads (with the arguments $\xxix$ on
the r.-h.\,side being suppressed)\,\footnote{Compare Eqs.(2.26) and (2.45) of  II  .} :
\be
Q^R_{ijkl}\xxix = - \bigl(\Gamma^R_{ik5}\,\Gamma^R_{jl5} - \Gamma^R_{jk5}\,\Gamma^R_{il5}\bigr)
\
\label{336}
\ee
leading, finally, to the total Lorentz indexed curvature tensor associated with the {\bf g$\prime$}-valued
part in \eqref{313} :
\be
R^R_{ijkl}\xxix= \bar R_{ijkl} + P_{ijkl}(x) + Q^R_{ijkl}\xxix .
\label{337}
\ee  
Here $\bar R_{ijkl}$ is the Riemannian part, $P_{ijkl}$ is the Weyl addition (compare App. A of Ref.\,\cite{2}), and 
$Q^R_{ijkl}$ is the tensor \eqref{336} originating from the de Sitter gauge aspect of the theory.
For the curvature scalar one computes from Eq.\eqref{336} with the
value $R_s$ for a $W_4$ as quoted in Eq.\eqref{230}, and with $\Gamma^R_{ij5}{=}-\Gamma^R_{ji5}$
having \ww $-\ha$ ,  
\be
R^R_s = R_s + Q^R ,
\label{338}
\ee
where $Q^R$ is given by
\be
Q^R = Q^R\xxix = \Gamma^R_{ij5}\xxix \,\,\Gamma^R{}^{ij5}\xxix .
\label{339}
\ee

We stress, however, that $Q^R$ is a scalar quantity characterizing the dynamics in the {\it fiber} of $\Et$ related
to what is called "the torsion of the Cartan connection" on $\widetilde P_W$. The space-time base is torsion-free
even in regions where $Q^R{\ne} 0$ as is seen from Eq.\eqref{212}. 
\vskip0.5cm

We now proceed in a similar manner as in Sect.\,2 above and introduce for the description of quark fields in the
$W_4$-standard model discussed here an enlarged spinor bundle $\St$ for hadrons analogous to the bundle
$\hat S$ for leptons in Eq,\eqref{221} : 
\be
\St = \St (W_4, \widetilde F{=}\C_4\times \widetilde{\C}\times R^+, \bar G_h) .
\label{340}
\ee
Here $\St$ is associated to the principal bundle $\bar{\widetilde P}_W(W_4,\bar G_h)$, with $\bar G_h$ denoting
the covering group of $G_h$ defined in (1.2) containing $Spin(4,1)$ instead of $SO(4,1)$ as a factor. $\widetilde\C$
in \eqref{340} is given by a product of a fiber $\hat\C$ for the assignments of left-handed (L) and right-handed (R)
Dirac spinor quark fields as in \eqref{221} above for the leptons times a 3-dimensional fiber $\C_3$ for the
colour-$SU(3)$ description of hadrons, i.e.
\be
\widetilde\C = \hat \C \times \C_3 \, .
\label{341}
\ee

The linearly transforming de Sitter covariant\,{\it chiral} fields $\pp_L(x)$ and $\pp_R(x)$ have to be 
constructed with the  projection operators
\be
P_{\pm}(\xix) = \ha \bigl(1\pm i \gafxix\bigl)
\label{341a}
\ee
with $+$ for the right-handed fields $(R)$ and $-$ for the left-handed fields $(L)$. Only the nonlinearly transforming
chiral Dirac - de Sitter spinor fields, $\pp_{L,R}\xxix$, (compare Eq.\eqref{328}), have the usual constant projection
factors $\ha (1\pm i\ga_5)$, being equal to $P_{\pm}(\xicirc)$, which appear also in the leptonic sector described
in Sect.\,2. We now formulate the gauge invariant hadronic Lagrangean using the linearly transforming fields,
$\pp_{L,R}(x)$, possessing as eigenvalues of $i \gafxix$ the values $-1$ and $+1$, respectively. We can go over to the
NL form, having the simpler $\ga_5$-structure, whenever necessary for clarity. We stress that this is a question of
representation and choice of gauge and does not restrict the de Sitter invariance of $\L_h$. 
Before writing down the Lagrangean density we restate again the
transformation properties of the chiral de Sitter spinor fields:
\be
\pp_L{}'(x) = [\si(x)]^{-\tf} S(A_{g(x)})\pp_L(x) = \ha \bigl(1-i\ga_5(\xixp)\bigl) \pp'(x)
\label{341b}
\ee
with $\xixp$ as given in Eq.\eqref{330} and $\pp'(x)$ according to Eq.\eqref{325} and analogously for $\pp_R{}'(x)$.

As mentioned, we consider in this paper only one generation of leptons and quarks
(compare Eqs.\eqref{222} and \eqref{223}). The chiral
Dirac spinor quark field sections, $\ppx$, of $\St$ appearing in $\L_h$ are:
\bea
\pp_L &=& {\pp^u_L\choose\pp^d_L} = {\ha\,(1-i\gafxi)\,\pp^u \choose\ha\,(1-i\gafxi)\,\pp^d} ,\quad Y{=}{1\over 6}
\label{342}  \\
\pp^u_R &=& \ha\,(1+i\gafxi)\,\pp^u , \qquad\qquad\qquad\hskip+2.0 mm Y{=}{2\over 3}
\label{343}  \\
\pp^d_R &=& \ha\,(1+i\gafxi)\,\pp^d , \qquad\qquad\qquad\hskip+2.0 mm Y{=}-{1\over 3}
\label{344} 
\eea
and their adjoints ($\ppb{=}\pp^\dagger \ga^0$) :
\bea
\ppb_L &=&  (\ppb{}^u_L,\ppb{}^d_L)=\bigl(\ppb{}^u\ha(1+i\gafxi),\,\ppb{}^d\ha(1+i\gafxi)\bigr),  Y{=}-{1\over 6}
\label{345} \\
\ppb{}^u_R &=& \ppb{}^u \,\ha(1-i\gafxi) , \qquad\qquad\qquad  Y{=}-{2\over 3}
\label{346} \\
\ppb{}^d_R &=& \ppb{}^d \,\ha(1-i\gafxi) , \qquad\qquad\qquad  Y{=}{1\over 3}
\label{347}
\eea

Taking due recognition of the Weyl weights $w(\tilde\xi){=}w(R){=}\ha$ and $w(\pp){=}-\tf$, the possible terms in
a $G_h$ gauge invariant, Hermitean,  
hadronic Lagrangean density $\L_h$ are (the arguments $(x)$ of the fields are suppressed) :
\bea
\L_h &=& K\sqrt {-g} \Biggl\{{1\over{12}}
\, g^{\mn} (\Dtr_\mu\xta)\,(\Dtr_\nu\xtb) {\eta_{ab}\over R^2}\, \phi^\dagger\phi 
\nonumber \\
& &+{i\over 2} \Bigl[(\ppb_L \gamx \orD\pp_L) + (\ppb{}^u_R \gamx \orD\pp^u_R) + (\ppb{}^d_R \gamx \orD\pp^d_R)
\nonumber \\
& &
 -( \ppb_L \olD \gamx\,\pp_L) -  (\ppb{}^u_R \olD \gamx \,\pp^u_R) - (\ppb{}^d_R \olD \gamx\, \pp^d_R) \Bigr]
\nonumber \\
& &
 + \l \Bigl[(\ppb_L\gaxi\,\pp_L) + (\ppb{}^u_R\gaxi\,\pp^u_R) + (\ppb{}^d_R\gaxi\,\pp^d_R)\Bigr]\cdot {\xta\over{R^2}}
\nonumber \\
& &
 + \bar\l R^2 \Bigl[\bigl(\ppb_L\ga^k(\tilde\xi)\,\pp_L\bigr)\bigl(\ppb_L\ga_k(\tilde\xi)\,\pp_L\bigr) + \bigl(\ppb{}^u_R
\ga^k(\tilde\xi)\,\pp^u_R\bigr)\bigl(\ppb{}^u_R\ga_k(\tilde\xi)\,\pp^u_R\bigr)
\nonumber \\
& &
\hskip5cm + \bigl(\ppb{}^d_R\ga^k(\tilde\xi)\,\pp^d_R\bigr)\bigl(\ppb{}^d_R\ga_k(\tilde\xi)\,\pp^d_R\bigr)\Bigr]
\nonumber \\
& &
 + \widetilde\l \Bigl[\bigl(\ppb_L\phi\bigr)\pp^u_R + \ppb{}^u_R\bigl(\phi^\dagger\pp_L\bigr) 
 + \bigl(\ppb_L\phi\bigr)\pp^d_R + \ppb{}^d_R\bigl(\phi^\dagger\pp_L\bigr) \Bigr]
\nonumber \\
& &
 -  {1\over 4}\,\delta_c\, G^s_{\mn}\,G_s^{\mn}\,\Biggr\} \, .
\label{348}
\eea

The first term in the curly brackets is the kinetic term for the de Sitter field $\xia$ multiplied by $\phi^\dagger\phi /R^2$
to give to this contribution the correct \ww and length dimension [$L^{-2}$]. Thereby a coupling between the square
of the modulus, $\Phi^2(x)$, of the $\phi$-field in the leptonic sector and the covariant 
derivative of the $\tilde\xi$-field is
appearing which, according to the discussion after Eq.\eqref{316} above, is only present in regions of the
space-time base where the de Sitter symmetry {\it does not reduce} to the Lorentz subsymmetry. 
This contribution is appearing here with the factor ${1\over{12}}$ since it plays a \rl  in $\L_h$ comparable to
the term $-{1\over{12}}R_s\phi^\dagger\phi$  in $\L_l$ of Eq.\eqref{229}. This choice of cofactor is adopted
in view of the later merging of the leptonic and hadronic sectors and the final use of the Lagrangean
$\L = \L_l + \L_h$ below. We remark in passing that the first term in the curly brackets is identical to
$-{1\over {12}}Q^R\phi^\dagger\phi$ for $D_\mu R(x){=}0$, i.e. after $D(1)$ symmetry breaking (compare
Eq.\eqref{338} and Eq.\eqref{48} below); hence a term of this type need not be separately introduced.

The following six terms in Eq.\eqref{348} represent the kinetic terms for the fermionic $(I{=}\ha ,L)$ 
and $(I{=}0,R)$ {\it up}\,$(u)$
and {\it down}\,$(d)$ massless quark fields, and the square brackets, multiplied by a 
dimensionless coupling constant $\l$, describe
the couplings between the $\xta$-field and the {\it up} and {\it down} quark fields. 
The Dirac matrices $\gamx$ are the $x$-dependent, nonlinearly under $SO(4,1)$
transforming, matrices of \ww $-\ha$ defined by $\ga^\mu(\xix) {=}\l^\mu_k(x)\,\ga^k(\xix)$ with the $\ga^k(\xix)$
transforming as in Eq.\eqref{333} having \ww $0$, 
and analogously for $\ga_a(\xix)$ appearing in the $\l$-term -- see footnote 23
implying that $\xia$ undergoes a $\Lambda(\xixp,\xix)$-rotation in the $\l$-term of \eqref{348} 
in changing the de Sitter gauge. There, due to the $LL$ and $RR$ structure of the chiral
quark currents, the contributions for $a{=}5$  in the summation over $a$ vanish due
to $i\gafxix\, P_{\pm}(\xix){=}\pm P_{\pm}(\xix)$ and $P_+(\xix)\cdot P_-(\xix)$ yielding zero. Stated differently,
one may express this property of the $\l$-term by saying that it would be identically vanishing for a 
zero-section on $\Et$, i.e. for $\xia\equiv\xiacirc=(0,0,0,0,-R(x))$.

$\orD$ and $\olD$ in Eq.\eqref{348} are total covariant differentiation operators acting,
respectively, to the right and to the left, differentiating every representation character the quantities
$\pp$  or $\ppb$ may
have (depending on the weak isospin) including de Sitter gauge covariant differentiation as well as
$SU(3)_c$ gauge covariant differentiation. The square brackets multiplied by the constant $\bar\l$ of dimension
$[L^{-2}]$ contain three current-current self-interaction terms of nonlinearly transforming Dirac spinor currents,
and the brackets multiplied by the constant $\widetilde\l$ of dimension $[L^{-1}]$ represent the analogon of the 
Yukawa-like term appearing in the leptonic Lagrangean \eqref{229} of Sect.\,2. 

We mention in passing that the current-current self-interaction terms proportional to $\bar\l$ allowed by the
Weyl weights may also contain each a Gell-Mann $SU(3)$ matrix $\l_s; s{=}1,2...8$\, 
between the $\ppb$ and $\pp$ fields with summation over $s$
in $\L_h$. This, however,  concerns the details of the colour gauge dynamics which clearly cannot be decided upon by
invariance arguments related to Weyl transformations alone. We thus have to leave this question of the
presence or absence of the $\l_s$-matrices in the current-current terms of Eq.\eqref{348} regarding the
chromodynamics as undecided.

Finally, the last term in the curly
brackets is the colour gauge curvature contribution with the summation over $s$ from 1 to 8 understood. The
analogous quadratic curvature invariants for the electroweak and $D(1)$ gauge groups appear already in $\L_l$
(see Eq.\eqref{229} above). 

We remark in passing that the length dimension of the curly brackets in Eq.\eqref{348} is $[L^{-2}]$, with the length dimensions of the quark fields being $[L^{-\ha}]$ as for the fermions in the leptonic sector in Sect.\,2 above. 

The contributions of the quark fields in Eq.\eqref{348} -- besides the $\widetilde\l$-terms --  could have been written shorter as
\bea
\sum_f\, \Bigl\{{i\over 2}[\ppb{}^f\,\big(\gamx\orD \,-\,\olD\gamx\bigr)\,\pp^f] + \l\,\bigl(\ppb{}^f \,\gaxi\,\, \pp^f\bigr)
\cdot{\xta\over{R^2}}    
\nonumber \\
+\,\bar\l\, R^2\,\bigl(\ppb{}^f\ga^k(\tilde\xi)\,\pp^f\bigr)\bigl(\ppb{}^f\ga_k(\tilde\xi)\,\pp^f\bigr) \bigr\}
\label{349}
\eea
with a summation over flavour $f$ (of which here only {\it up} and {\it down} is considered). However, we prefer the explicit form given in \eqref{348} in parallel to the form \eqref{229} for the leptons, since its $\ga_5$-structure is more 
transparent. 

\newpage

% Beginn Kapitel IV=====================================================

\section{Field Equations\\ and Weyl Symmetry Breaking}

\qquad The field equations following from the $G_l$ gauge invariant leptonic Lagrangean \eqref{229} were
discussed in detail in  I  . Before breaking the Weyl invariance and going over to a formulation of the standard model 
in a realistic Riemannian description for the classical gravitational field, we derive in this section the field equations
following from the $G_h$ gauge invariant Lagrangean \eqref{348}. Those additional contributions, which arise
when one goes over to the total Lagrangean $\L{=}\L_l+\L_h$ combining both sectors and discuss its $D(1)$ 
symmetry breaking, will be investigated at the end of the section.  

Varying the quark fields $\pp_L$, $\pp^u_R$ and $\pp^d_R$ in \eqref{348} one derives the following
three Dirac-type equations each of which possessing an additional cubic self-interaction term proportional to $\bar\l$ :
\bea
i\ga^k(\xix)\,D^R_k\pp_L(x)&+&\l\,{1\over{R^2(x)}}\, \ga_a(\xix)\,\xia\,\pp_L(x) 
\nonumber \\
&+&\widetilde\l\,\phi(x)\bigl[\pp^u_R(x)+\pp^d_R(x)\bigr]
\nonumber \\
+2\,\bar\l\,R^2(x)& &\hskip-10mm\Bigl(\ppb_L(x)\,\ga_k(\xix)\,\pp_L(x)\Bigr)\,\ga^k(\xix)\,\pp_L(x)= 0 ,
\label{41}
\eea
\bea
i\ga^k(\xix)\,D^R_k\pp^u_R(x) &+&\l\,{1\over{R^2(x)}}\,\ga_a(\xix)\,\xia\,\pp^u_R(x)
\nonumber \\
&+&\widetilde\l\,\Bigl(\phi^\dagger(x)\pp_L(x)\Bigr)
\nonumber \\
+2\,\bar\l\,R^2(x)& &\hskip-10mm\Bigl(\ppb{}^u_R(x)\,\ga_k(\xix)\,\pp^u_R(x)\Bigr)\,\ga^k(\xix)\,\pp^u_R(x)=0 ,
\label{42}
\eea
\bea
i\ga^k(\xix)\,D^R_k\pp^d_R(x)&+&\l\,{1\over{R^2(x)}}\,\ga_a(\xix)\,\xia\,\pp^d_R(x)
\nonumber \\
&+&\widetilde\l\,\Bigl(\phi^\dagger(x)\pp_L(x)\Bigr)
\nonumber \\
+2\,\bar\l\,R^2(x)& &\hskip-10mm\Bigl(\ppb{}^d_R(x)\,\ga_k(\xix)\,\pp^d_R(x)\Bigr)\,\ga^k(\xix)\,\pp^d_R(x)=0 .
\label{43}
\eea

In each of these equations $\widetilde\l$ multiplies a term coupling right-handed and left-handed fields like in the
term proportional to $\tilde\ga$ in $\L_l$ of Eq.\eqref{229} and the field equations derived from it in  I  describing
electrons and electron neutrinos. The $\l$-terms in the equations \eqref{41} - \eqref{43} -- when taken alone -- could
be interpreted as mass-like terms for the $\pp$-fields depending on $\xia$ and $R(x)$ which simplify
after $D(1)$ symmetry breaking as will
be shown below. The $\pp\pp\pp$-terms obtained from the current-current self-interaction proportional to $\bar\l$
in \eqref{348} are evidently also related to the effective masses of the quark fields if at all one insists in introducing
the notion of a mass for the $\pp$-fields. We shall see below that the quark currents and quark fields are
confined to small domains in space-time while the concept of a mass of a field or particle applies in the strict sense
only to fields or particles appearing asymptotically.

Varying  now the de Sitter fields, $\xia$, in Eq.\eqref{348} one obtains, with $\phi^\dagger(x)\phi(x)\,{=}\Phi^2(x)$ 
and $\gaxi{=}\ga_a(\tilde\xi(x))$ (see also App.\,B):
\bea
\six\, \Dtr{}^\mu \Dtr_\mu\, \tilde\xi_a(x) - {2\over R(x)}\bigl(\Dtr{}^\mu\tilde\xi_a(x)\bigr) \widetilde D_\mu R(x) + 
\bigl(\Dtr{}^\mu\tilde\xi_a(x)\bigr)\, {1\over\Phi^2(x)}\,D_\mu \Phi^2(x)  
 \nonumber \\
 - {\l\over\Phi^2(x)} \Bigl[\bigl(\ppb_L(x) \gaxi\, \pp_L(x)\bigr)+\bigl(\ppb{}^u_R(x) \gaxi\, \pp^u_R(x)\bigr)+
   \bigl(\ppb{}^d_R(x) \gaxi \,\pp^d_R(x)\bigr)\Bigr]
\nonumber \\
 - {R^2(x)\over{\Phi^2(x)}}\cdot \delta '\L_h=0 .
\label{44}
\hskip+4cm
\eea
The quark current term arising from the explicit $\tilde\xi(x)$-dependence of the $\l$ term in Eq.\eqref{348} acts here as a 
source for the $\xia$-field. The last term on the l.-h.s. of Eq.\eqref{44} is meant to represent those contributions
originating from a possible implicit $\tilde\xi(x)$-dependence of the $\ga_a$-matrices in \eqref{348} -- and in particular of
the projection operators \eqref{341a} contained in the chiral spinor fields. Such contributions arising from the
variation of the $\xia$ in $\L_h$ have to be discriminated from gauge degrees of freedom and are investigated
separately in Appendix B. There it is shown that the $\tilde\xi(x)$-dependence of the $\ga^a$-matrices in \eqref{348}
is a gauge effect being dynamically irrelevant in a variational derivation of the field equations for $\xia$. Hence
the term $\delta '\L_h$ in Eq.\eqref{44} is zero and can be ignored.

Eq.\eqref{44} represents a complicated set of equations for $\tilde\xi_a(x); a{=}0,1,2,3,5,$ which we only need and therefore discuss here in the broken Weyl case, i.e.\,{\it after} breaking the $D(1)$ symmetry by 
demanding, as was shown and discussed
in detail in  I , for the scalar isospinor field $\phi(x)$ of Sect.\,2 the symmetry breaking condition 
\be
D_\mu \Phi^2(x) \equiv \pa_\mu\Phi^2(x) + \ka_\mu\Phi^2(x) = 0 ,
\label{45}
\ee
freezing the norm $\Phi(x)$ of $\phi(x)$ to a constant $x$-independent value called $\hph$ in  I  . We shall discuss
the Weyl symmetry breaking in more detail below in this section and only mention at this place that, with Eq.\eqref{45},
also the variable curvature radius, $R(x)$, of the local de Sitter fibers in $\Et$ have to obey an analogous equation 
on the hadronic side, i.e. $\widetilde D_\mu R(x){=}0$, freezing the radius $R(x)$ to a constant $x$-independent 
value $R_0$. After $D(1)$ symmetry breaking the Eqs.\eqref{44} thus take the much simpler form with
$\ga_a(\xi){=}\ga_a(\xi(x))$ :
\bea
\six\,D^{R\,\mu} D^R_\mu\,\xi_a(x) = \hskip+8cm
\nonumber \\ 
{\l\over\hph{}^2} \Bigl[\bigl(\ppb_L(x)\ga_a(\xi)\,\pp_L(x)\bigr) +
 \bigl(\ppb{}^u_R(x)\ga_a(\xi)\,
 \pp^u_R(x)\bigr) + \bigl(\ppb{}^d_R(x)\ga_a(\xi)\,\pp^d_R(x)\bigr)\Bigr] ,
\nonumber \\
\label{46}
\eea 
where we have left out the tildes on the $\xi$-fields and the covariant derivatives since the Weyl vector field $\ka_\mu$
is zero after Weyl symmetry breaking (see below) and the bundle $\Et$ of Eq.\eqref{32} raised over a $W_4$ reduces
to the bundle $E$ of Eq.\eqref{31} raised over a $V_4$ with $W_4 \longrightarrow V_4$ in the Weyl symmetry
breaking limit. Eq.\eqref{46} shows that the local de Sitter quark field current (here for the first generation of quarks)
between the square brackets on the r.-h.s., multiplied by the dimensionless constant $\l'{=}\l /\hph{}^2$, acts
as the source current for the $\xi_a(x)$-field obeying $\xi^a(x)\xi_a(x){=}-R_0^2$. We, moreover, observe that in the equations \eqref{44} and \eqref{46}  the quark current is zero for $a{=}5$ due to the chirality of the fermion fields as a result
of $P_L\cdot P_R$ yielding zero, as was already noted after Eq.\eqref{348} above. Furthermore, for 
$D^R\xi^a(x){=}0$ , i.e. when according to Eq.\eqref{316} the de Sitter gauge symmetry reduces to the Lorentz subsymmetry describing pure gravitation, the Eqs.\eqref{46} imply for $a{=}k{=}0,1,2,3,$ that the total
Lorentz current of the $L$ and $R$ quark fields has to vanish. This means that the r.-h.s. of Eq.\eqref{46} 
can only be nonvanishing for those regions in space-time where $D^R_\mu\xi^a(x)\equiv V^a_\mu(x)\not=0$.  
Outside these regions the total first generation quark current of NL type with Lorentz components, $k{=}0,1,2,3,$ 
appearing in the square brackets of Eq.\eqref{46}, {\it have to vanish}. This property expresses the idea of confinement,
 formulated in the present context and after $D(1)$ symmetry breaking, making use of the section $\xi^a(x)$
defined on the de Sitter bundle $E$ playing here the \role of a {\it symmetry reducing field associated with
strong interactions}: Only gravitation can appear asymptotically, not, however, the quark currents and quark fields.

To underline this result we contract Eq.\eqref{46} with the fields $\xi^a(x)$ yielding in the broken case, with
$R(x){=}R_0{=}const$ : 
\bea
\six\,V^\mu_a(x)\,V^a_\mu(x) =   
 -\l ' \Bigl[\bigl(\ppb_L(x)\ga_a(\xi(x))\pp_L(x)\bigr) +  \bigl(\ppb{}^u_R(x)\ga_a(\xi(x))\pp^u_R(x)\bigr) +
\nonumber  \\
\bigl(\ppb{}^d_R(x)\ga_a(\xi(x)) \pp^d_R(x)\bigr)\Bigr]\cdot \xi^a(x)  ,\qquad\qquad
\label{47}
\eea
where, as mentioned above, on the r.-h.s. in the summations over $a$ only $a{=}k{=}0,1,2,3$ can contribute.
We evaluate the invariant on the l.-h.s. in the NL form using, after $D(1)$ symmetry breaking, for those regions
of space-time where the de Sitter symmetry does {\it not} reduce (compare the discussion after Eq.\eqref{316} above)
\be
\buildrel{NL}\over V{}\hskip-2.0 mm^a_\mu(x) \equiv 
\buildrel{NL}\over{\widetilde D}{}\hskip-2.0mm^R_\mu\xiacirc =\left \{
\begin{array}{l}
0\,\quad\hskip3.06cm{\rm for}\, a{=}5 ,  \\
 R_0 \,\Ga^R{}_\mu{}^i{}_5(x,\xi(x) \,\qquad{\rm for}\,a{=}i{=}0,1,2,3,   \\
\end{array}\right.
\label{48}
\ee
and find the interesting result that
\bea
V^\mu_a(x)\,V^a_\mu(x) = \buildrel{NL}\over V{}\hskip-2.0 mm^a_\mu(x)\buildrel{NL}\over V{}\hskip-2.0 mm^\mu_a(x)
&=& - R_0{}^2\, \Ga^R_{ij5}(x,\xi(x))\, \Ga^{R\,\,{ij5}}(x,\xi(x))  
\nonumber  \\
&=& - R_0{}^2\, Q^R(x,\xi(x)) ,
\eea
where in the last step Eq.\eqref{339} has been used. This leads, finally, to the following form of Eq.\eqref{47} :
\bea
\six\,Q^R(x,\xi(x)) = {\l '\over{R^2_0}}\Bigl[\bigl(\ppb_L(x)\,\ga_a(\xi(x))\,\pp_L(x)\bigr)+\bigl(\ppb{}^u_R(x)\,\ga_a(\xi(x))\,\pp^u_R(x)\bigr)
\nonumber  \\
+ \bigl(\ppb{}^d_R(x)\,\ga_a(\xi(x))\,\pp^d_R(x)\bigr)\Bigr]\cdot\xi^a(x) . \qquad\qquad
\label{410}
\eea
We interprete Eq.\eqref{410} by stating that those regions of space-time, where the de Sitter symmetry 
{\it does not reduce} to the Lorentz subsymmetry, there exists a proportionality between the total Dirac-de Sitter
covariant [NL] quark current, contracted with the field $\xi^a(x)$, and the scalar $Q^R$ 
derived from the torsion of the Cartan connection on the bundle $\widetilde P_W$ over $W_4$. 
Thus in locally confined regions where quark currents do exist the geometry is characterized by 
de Sitter induced torsion. Stated differently this implies, that the regions in space-time
where strong interaction takes place, are geometrically to be characterized by the presence of a nonvanishing $Q^R$.

After this discussion and in going over in Eq.\eqref{41} to the nonlinearly transforming spinor fields\footnote{$S^{-1}
\bigl(A(\xi(x)\bigr)\,\pp_L(x)=\ha(1-i\ga_5)\,\pp(x,\xi(x))\equiv\pp_L(x,\xi(x))$}, $\pp_L(x,\xi(x))$,
introduced in Eq.\eqref{328} -- with $\xia$ going over, as mentioned, into $\xi^a(x)$ defined on $E$ transforming
as in Eq.\eqref{34} and obeying $\xi^a(x)\xi_a(x){=}-R_0{}^2$, with $x\in V_4$, one can rewrite this
equation, using Eqs.\eqref{331} and \eqref{332} in the $\l$-term,  in the form
\bea
i\ga^k\buildrel{NL}\over D_k \pp_L(x,\xi(x)) &-& \l \,{{\ga_5(\xi(x))}\over {R_0}}\,\pp_L(x,\xi(x)) 
\nonumber \\
&+&\widetilde\l\,\phi(x)\bigl[\pp^u_R(x,\xi(x))+\pp^d_R(x,\xi(x))\bigr]
\nonumber \\
+2\,\bar\l\,R^2_0& &\hskip-10mm\Bigl(\ppb_L(x,\xi(x))\,\ga_k\,\pp_L(x,\xi(x))\Bigr)\,\ga^k\,\pp_L(x,\xi(x)) =0
\nonumber \\
\label{411}
\eea
together with two similar equations for $\pp^u_R$ and $\pp^d_R$ following
from \eqref{42} and \eqref{43}, respectively. 
Here $\ga^k$ are the constant Dirac matrices of Eq.\eqref{323}, and $\ga_5(\xi(x)){=}-\ga_a\,\xi^a(x)/{R_0}$, with
the "mass-like term" proportional to $\l$ being scaled by the fixed fiber radius $R_0$. 

We also note that the $U(1)_Y\times SU(2)_W\times SU(3)_c$
gauge covariance is not affected by the transition from Eq.\eqref{41} to Eq.\eqref{411}.
In order to study further the Eq.\eqref{411} and its two coupled companions following from \eqref{42} and
\eqref{43} one has to go over to the electromagnetic gauge described in  I  in which $\phi(x)$ takes the form
$\hat\phi=\left({0\atop \hph}\right)$ with constant $\hph$ after $D(1)$ symmetry breaking (compare App.\,A
and B of  I  ). We shall, however, not pursue this question further here and return to our main topic.

Before we turn to the problem of Weyl symmetry breaking in detail we quote the field equation for the scalar
isospinor field $\phi(x)$ following from $\L=\L_l+\L_h$ :
\bea
\delta\,\phi^\dagger :\quad g^{\mn}\,D_\mu D_\nu\, \phi \hskip-2.0mm&+&\hskip-2.0mm
\six R_s\,\phi + 4\beta(\phi^\dagger\phi)\phi -2\tilde\ga
\,\pbr\,\psi_L \nonumber \\
+\six Q^R \phi \hskip-2.0mm&+&\hskip-2.0mm\six {1\over{R^2}}
(D^\mu R)(D_\mu R)\phi -2\tilde\l\, (\ppb^u_R\,\pp_L + \ppb^d_R\,\pp_L )=0 .
\nonumber  \\
\label{411a}
\eea
Here the first line of Eq.\eqref{411a}  is due to $\L_l$ (compare Eq.(2.35) of  I  ) and the second line is 
derived from $\L_h$. The term $\six Q^R\phi$ and the following one originate from an expression similar
to the one given in Eq.\eqref{48} prior to $D(1)$ symmetry breaking. We observe that the $Q^R$ term in the 
second line combines with the corresponding term in the first line to yield a contribution $\six R^R_s \phi$
in the $\phi$-equation involving the total curvature scalar $R^R_s {=}R_s{+}Q^R$. Outside the domains of
strong interactions the $Q^ R$ term is zero as is the $\tilde\l$-term involving the quark fields. Moreover, the
$(D^\mu R)(D_\mu R)$ term in the second line of \eqref{411a} disappears in the $D(1)$ symmetry breaking limit,
so we reobtain the same framework for $\phi$ as discussed in  I for regions where $\Dtr_\mu\xia{=}0$. 

For the gauge covariant symmetric energy-momentum tensor of the $\phi$-field one now obtains 
by varying the metric in $\L_l + \L_h$  (correcting a sign error in Eq.(2.46) of  I  ) :
\bea
\Theta^{(\phi)}_{\mn} &=& \ha\,\bigl[(D_\mu\phi)^\dagger\,(D_\nu\phi) + (D_\nu\phi)^\dagger\,(D_\mu\phi)\bigr]
- \six\bigl\{D_{(\mu}D_{\nu)}\Phi^2 - g_{\mn}\,D^\r D_\r \Phi^2\bigr\}
\nonumber \\
 &-& g_{\mn}\,\bigl[\ha\,g^{\r\l}(D_\r\phi)^\dagger(D_\l\phi) - \beta\,(\phi^\dagger\phi)^2\bigr]
\nonumber \\
&+&\six\bigl[(\Dtr_\mu\xta)(\Dtr_\nu\xtb) - \ha\,g_{\mn}\,(\Dtr{}^\r\xta)(\Dtr_\r\xtb)\bigr]\, {\eta_{ab}\over{R^2}} 
\,\Phi^2 
\label{411b}
\eea
The first three terms of \eqref{411b} originate from $\L_L$, the last line is due to $\L_h$. This latter part vanishes 
again outside the regions of strong interactions and \eqref{411b} reduces to the expression given in  I. The addition
in \eqref{411b} derived from $\L_h$ may be expressed in terms of $\Gamma^R_{ik5}\xxix$ yielding the result : 
\be
\six\Bigl[\Gamma^R_{\mu i5}\,\Gamma^R_{\nu k5}\, \eta^{ik} - {{1}\over{R^2}}( D_\mu R)\,(D_\nu R)
 + \ha\,g_{\mn}\bigl(Q^R + {{1}\over{R^2}} (D^\r R)\,(D_\r R)\bigr)\Bigr]\,\Phi^2 .
\label{411c}
\ee
Here again the terms formed with $D_\r R$ disappear in the $D(1)$ symmetry breaking limit.  

\vskip5.0mm
We now come back to the details of the Weyl symmetry breaking leading from the established Weyl invariant
and scaleless theory given by the total Lagrangean $\L{=}\L_l {+} \L_h$ defined over a Weyl space-time, $W_4$,
to a formulation of the standard model in a Riemannian space-time, $V_4$, including gravitation in the form of
Einstein's theory of general relativity, together with the introduction of a definite and universal unit of mass
or length into the theory as the result of the $D(1)$ symmetry breaking.

To this end we follow closely the procedure presented in  I  by adding a $D(1)$ symmetry breaking term $\L_B$ of
\ww $1$ to the total Lagrangean $\L$ being constructed with the help of the $U(1)_Y\times SU(2)_W$ invariant
squared modulus of the scalar isospinor field $\phi$ and the curvature scalar $R_s$ of the ambient Weyl space :  
\be
\L_B = - {a\over 2}\,K\sqrt{-g}\, \bigl\{{1\over 6} R_s + \big[{mc\over\hbar}\bigr]^2\,\phi^\dagger\phi\bigr\} .
\label{412}
\ee
$\phi$ is defined as a section on the bundle $E'$ introduced in Eq.\eqref{226} with $\phi^\dagger\phi{=}\Phi^2$
multiplied in \eqref{412} by the squared inverse length, $[mc/\hbar]^2$, which is the universal length scale
to be introduced into the theory being associated with the mass $m$ of the scalar field.

The idea behind the form of $\L_B$ as given in \eqref{412} is that the masses in the theory arise from
an interplay of a geometric quantity, the curvature scalar $R_s$ of the ambient Weyl space,  and a field
quantity, $\Phi^2$, associated with the scalar isospinor field $\phi$ with $m$ being the mass of this field in the
broken case.  In Ref.\,\cite{2} this kind of mass generation from a geometric quantity --- the curvature scalar
of the underlying space-time geometry --- and a particular field quantity --- the norm squared of $\phi$ --- 
with $\phi$  representing a characteristic object, having a mass and  existing 
within this geometry, was referred to as a kind of 'Archimedes principle' . The surprising result
in Refs.\,\cite{2} and \cite{1} was that this symmetry breaking following, as far as physics and the energy density
is concerned, from adding the Lagrangean $\L_B$ to $\L_l$, and as far as mathematics is concerned, from requiring 
the condition $D_\mu\Phi^2{=}0$, lead to the consequence that also 
other, previouly massless, field quantities -- the weak bosons $W^\pm$ and $Z^0$ as well as the
electron -- became massive fields being measured in units of $m$ (compare Eqs. (3.52) and (3.55) of  I  ). 
As shown there, these new mass terms appeared via the energy-momentum tensor $\Theta^{(\phi)}_{\mn}$
of the $\phi$-field. 
Combining the leptonic and the hadronic sector
we now base the breaking of the Weyl symmetry on the total Lagrangean $\L_l+\L_h+\L_B$ together with
the condition \eqref{45}, the latter leading immediately to $\kappa_\mu{=}-\partial\log\Phi^2$ and the integrable 
Weyl case with $f_{\mn}{=}0$.

The constant $a$ appearing in Eq.\eqref{412} was treated in  I  as a Lagrangean multiplyer and as a means to
track the new terms in the equations derived after symmetry breaking. Ultimately, in the field equations
and in the equations derived from them and from the constraint following from $\L_B$ in varying $a$
(see Eq.(3.11) of  I  ), $a$ is set equal to 1. Similarly we can now proceed in the present context,
however, we shall not work out the consequences in full detail here.

Since the symmetry reduction investigated here is governed by a theorem of differential geometry which was
referred to above \footnote{See footnote 21 above.} and described in detail at the end of Sect.\,2 and in Sect.\,3
of  I\,, we here only repeat that the $D(1)$ symmetry is explicitly broken by the squared modulus of the scalar 
isospinor field $\phi(x)$ in demanding the symmetry breaking relation \eqref{45}. The squared modulus, $\Phi^2$,
is thus the actual vehicle for $D(1)$ symmetry breaking leading, as was show in Sect.\,3 of  I, in the broken case
including $\L_B$ and in using also the field equations, to the complete reduction of the
Weyl space $W_4$ to a Riemann space $V_4$ yielding finally       
\be
\ka_\mu = 0 \qquad {\rm and}\qquad \Phi^2  = const = \hph{}^2
\label{413}
\ee
together with the established length\, $\hbar/mc$ \,as the universal scale of lengths in the theory.
At the same time the mass terms of the weak bosons and of the electron
(expressed in unit of $m$) appear on the scene.

In this process of Weyl symmetry breaking the electroweak gauge group $\widetilde G{=}U(1)_Y\times U(2)_W$ is
{\it not} broken but is nonlinearly realized on the stability subgroup $SU(1)_{e.m.}$ being the electromagnetic 
subgroup of $\widetilde G$ (see App.\,A and B of  I ).

This situation in the leptonic sector is completely analogous to the symmetry reduction in the hadronic sector
governed by Eqs.\eqref{313} and \eqref{316}, i.e. by $\Dtr\xia{=}0$, from which follows by contraction with
$\tilde\xi_a(x)$ immediately (compare also \eqref{319} above)\,\footnote{We have left out here the label $R$ 
on the Weyl covariant derivative of $R(x){}^2$ since the de Sitter connection is no longer involved.} 
\be
\Dtr_\mu\bigl(\xia\tilde\xi_a(x)\bigr) = - \widetilde D_\mu R(x){}^2 = 0
\label{414}
\ee
which is analogous to Eq.\eqref{45} for $\Phi^2(x)$. Thus the section $\xia$ on $\Et$ is the symmetry reducing field
governing the transition $SO(4,1)\longrightarrow SO(3,1)$ in the hadronic sector as studied in Sect.\,3 above, with 
the full de Sitter symmetry being realized only in those regions of space-time where the local quark currents of the
r.-h.s. of Eq.\eqref{46} or \eqref{410} persist, but this $SO(4,1)$ symmetry is realized {\it nonlinearly} on the
$SO(3,1)$ subgroup as expressed by the Dirac-type equations for the quark fields (see Eqs.\eqref{41} and \eqref{411}). 
Asymptotically, i.e. far away from the strong interaction sources given by the quark currents, the $SO(4,1)$ gauge
symmetry is broken down to a  linearly realized Lorentz subsymmetry describing pure gravitation as in general
relativity.

Having broken the Weyl symmetry by \eqref{45} together with \eqref{412} yielding \eqref{413} the consequence of
Eq.\eqref{413} is that the curvature radius of the fiber of the soldered bundle $\Et$ is a constant, i.e. $R(x)\equiv R_0$,
 and $\Et$ reduces to $E$ with the Riemannian base $V_4$ -- as was already incorporated into Eq.\eqref{411} above. 

The final step which has to be carried out now is the proper disentanglement of the neutral and charged
components in the quark sector and the working out of the details of the electroweak interaction in a similar manner 
as it was done for the leptonic sector in  I.
This investigation is necessary in order to determine those hadronic source current components coupling to
electromagnetism, being thus 'visible' from outside, and discriminate them from the neutral components. This
analysis is more complicated due to the nonlinear current-current interaction present in the hadronic Lagrangean
$\L_h$. We have to come back to this problem in a separate investigation.

%Beginn Kapitel V==========================================================

\section{Concluding Remarks}

\qquad  In this paper we have extended the electroweak theory for leptons studied in Refs.\,\cite{1} and \cite{2}
to a complete formulation of the standard model basing it on a Weyl invariant Lagrangean defined over a Weyl
space-time $W_4$ containing only massless fields. At the same time we extended in the hadronic sector of the
theory the Lorentz group, as gauge group of gravity in a vierbein formulation, to the de Sitter group and 
a de Sitter gauge theory related 
to a geometric formulation for a theory of strong interactions based on the gauge group $SU(1)_Y {\times} SU(2)_W
{\times} SU(3)_c$ incorporating gravitation in a unified space-time description by considering for strong
interacting fields also an  internal geometric arena --- here a fiber of a bundle $\Et$ over space-time --- possessing
a homogeneous space of the $SO(4,1)$ de Sitter group as fiber. The radius of curvature of the fiber was treated as
a gauged Weyl degree of freedom. The section $\xia$ on the de Sitter fiber bundle $\Et$ played the \rl of a 
Higgs-type field for strong interaction associated with the symmetry breaking $SO(4,1) \longrightarrow SO(3,1)$
guaranteeing -- together with the $D(1)$ symmetry breaking in the leptonic sector -- that far away from the material
sources of the geometry only light and gravity is observable.

One of the essential criteria besides spin and isospin in determining the contributions appearing in the hadronic
Lagrangean is the \ww of the fields in question. It was found that a four fermion current-current self-interaction
for the 4-component quark spinor fields $\pp$ was allowed by the \ww of these fields, which gives to the discussion
of the effective masses of quarks a new and different meaning. It was, furthermore, found that confinement of quark
fields and quark currents is indeed observed in this de Sitter framework and is characterized by de Sitter induced
torsion, i.e. by the torsion of the Cartan connection governing the geometry of the de Sitter gauge 
theory in those regions of space-time where strong interactions exist. These areas are geometrically characterized
by the addition of a quantity $Q^R$ to the curvature scalar which is expressible as an invariant constructed from
the $(i,5)$ components of the de Sitter gauge curvature representing the so called torsion of the Cartan connection
on the frame bundle $\widetilde P_W$ to which $\Et$ is associated.

Then, in a final step, the Weyl symmetry is broken explicitly by demanding that the scalar isospinor field
$\phi$ possesses a norm which is Weyl covariant constant (see Eq.\eqref{45}). 
This changes the $\widetilde G{=}SU(1)_Y{\times}SU(2)_W$
electroweak symmetry transformations, as described in detail in I, to the nonlinear realization of the transformations
of $\widetilde G$ on the electromagnetic subgroup $U(1)_{e.m.}$ of $\widetilde G$, without, however, reducing the
$\widetilde G$-gauge symmetry completely to the electromagnetic sub-symmetry --- a phenomenon called
'spontaneous symmetry breaking' in the conventional treatment. At the level of the Lagrangean the Weyl
symmetry breaking is induced by adding the expression $\L_B$ of Eq.\eqref{412} to the total Lagrangean $\L$.
Thereby the Weyl symmetry is broken with the $W_4$ going over into a $V_4$ of general relativity, and the
Compton wave length of the field $\phi$ is established as a universal length scale in the theory. In this process
the norm $\Phi$ of the field $\phi$ is reduced to a constant $\hph$. The constant $\hph^2$ enters the 
gravitational constant in Einstein's equations for the metric in a Brans-Dicke like manner as was shown and 
discussed in detail in  I. This is true unchanged in the present context except for additional contributions to
the energy-momentum tensors for quark matter involving the spinor fields $\pp_L$, $\pp^u_R$ and $\pp^d_R$
in analogy to Eq.\,(2.47) of I, as well as the colour gauge curvature fields $G^s_{\mn}$ entering $T^{(G_s)}_{\mn}$
defined in analogy to Eqs.\,(2.48)--(2.50) in  I.

As an additional consequence after $D(1)$ symmetry breaking with $W_4 \longrightarrow V_4$ one should mention the 
freezing of the variable radius of curvature $R(x)$ of the de Sitter fiber of $\Et$ to a constant $R_0$ the value
of which, however, cannot be determined from the Weyl symmetry breaking mechanism alone.

\vspace{0.5cm}

%====================================================================
%\newpage
%APPENDIX
\begin{appendix}
\vskip0.00cm
\section{Appendix A}

\qquad In this appendix we assemble some formulae concerning the Clifford algebra of the $4\times4$ Dirac matrices
used obove. Some of the equations of the text are repeated for easier reading. The notation is the same as 
in the main text, however, the $D(1)$ degree of freedom scaling the fiber radius is disregarded. $G{=}SO(4,1)
{=}O(4,1)^{++}$ and $H{=}SO(3,1){=}O(3,1)^{++}$, with $++$ denoting the special orthochronous groups.
\bea
&&\{\ga_a,\ga_b\} = 2\, \eta_{ab}\,\cdot {\bf 1}\, ;\quad \eta_{ab}{=}diag(1,-1,-1,-1,-1)  ,\quad  a,b{=}0,1,2,3,5,
\nonumber \\
&&\{\ga_i,\ga_k\} = 2 \, \eta_{ik} \, \cdot {\bf 1}\, ; \quad\eta_{ik}{=}diag(1,-1,-1,-1)  , \quad\qquad    i,k{=}0,1,2,3,
\nonumber \\
\ga^a {=}&&\hskip-6mm \bigl(\ga^k;k{=}0,1,2,3,\, \ga^5{=}\ga^0\ga^1\ga^2\ga^3\bigr);\quad \ga^a{=}\eta^{ab}\ga_b,\quad\ga^a{}^\dagger
{=}\ga^0\ga^a\ga^0, \quad (\ga^5)^2{=}-1.
\nonumber
\eea

The Clifford algebra of the $4\times 4$ matrices $\ga^k$ has $1+4+6+4+1{=}16$ elements conventionally
denoted by $S,V,T,A,P$ :
\bea
&&\quad {\bf 1},\quad \ga^k,\quad\ga^i\,\ga^k ;{\footnotesize i<k},\quad \ga^i\,\ga^5, \quad\, \ga^5
\nonumber \\
&&\quad S\quad\, \, V \quad\qquad\quad T\quad \quad\qquad A \qquad P 
\nonumber
\eea
\hskip4.4mm The chirality operators are constructed with ${\bf 1}$ and $i\ga_5$ : $P_{\pm}{=}\ha(1\pm i\ga_5)$.
The Lorentz generators, i.e. the 6 generators of $Spin(3,1)$ are : $S^{ik}{=}{i\over 4}[\ga^i,\ga^k]$,
the de Sitter generators, i.e. the 6+4 generators of $Spin(4,1)$ are: $S^{ab}{=}{i\over 4}[\ga^a,\ga^b]$,
with the de Sitter spinor boost generators being $S^{i5}{=}{i\over 4}[\ga^i,\ga^5]$.

With the help of the boost transformation with the $5\times 5$ matrix
\be
\bigl[A(\xix)\bigr]^a{}_b =\left(\begin{array}{cc} \delta^i_j+{{\tilde\xi^i(x)\tilde\xi_j(x)}\over{R(x)(R(x)-\tilde\xi^5(x))}}
& -{{\tilde\xi^i(x)}\over{R(x)}} \cr
-{{\tilde\xi_j(x)}\over{R(x)}} & -{{\tilde\xi^5(x)}\over{R(x)}} \cr
\end{array}\right)
\label{A1}
\ee
sending $\xiacirc{=}(0,0,0,0,-R(x))$ into $\xia ; a{=}0,1,2,3,5,$ i.e. transforming the point $\xicirc$ of the local
fiber isomorphic to $G/H$  --  with $\xicirc$ possessing the subgroup $H$ as isotropy group -- into the point
$\xix{\in}F_x\simeq G/H$, we now go over, with the help of the associated spinor boost $S\bigl(A(\xix)\bigr)$ 
given below, to a new representation of the $\ga^a$-matrices and chiral projection
operators associated with the point $\xix$ on the fiber $F_x$ of $\Et$ [compare Eqs.\eqref{331} and \eqref{341a}]: 
\bea
\ga^a &&\longrightarrow\quad \ga^a(\xix) = \bigl(\ga^k(\xix),\ga^5(\xix)\bigr) ,
\label{A2} \\
P_{\pm}{=}\ha (1\pm i\ga_5)&&\longrightarrow\quad P_{\pm}(\xix){=}\ha \bigl(1\pm i\ga_5(\xix)\bigr) .
\label{A3} 
\eea

The soldering properties of the bundle $\Et$ at the contact point $\xicirc$ remain unaffected. The same is true for
 the representation of the generators of the group $Spin(3,1)$ associated with the point $\xicirc$ as well as the generators
of $Spin(4,1)$ related to the transformations in the embedding space $R_{1,4}$.

While the original representation of the $\ga^a$-matrices defines the homomorphism $Spin(4,1)\longrightarrow SO(4,1)$
through the relation \eqref{327}, i.e.
\be
S(A_{g(x)})\,\ga^a\,S^{-1}(A_{g(x)}) =\bigl [A^{-1}_{g(x)}\bigr]^a{}_b\, \ga^b ,
\label{A4}
\ee
the new representation defines the same homomorphism through the relation (see footnote 23)
\be
S(A_{g(x)})\,\ga^a(\xix)\,S^{-1}(A_{g(x)}) = \bigl[A^{-1}\bigl(\Lambda (\xixp,\xix)\bigr)\bigr]^a{}_b \,\ga^b(\xixp) .
\label{A5}
\ee
We called the $\ga^a(\xix)$ the under $SO(4,1)$ {\it nonlinearly} transforming $\ga^a$-matrices (abbreviated by
$[NL]$ in Sects.\,3 and 4).
 Eq.\eqref{A5} follows
from Eq.\eqref{A4} with the help of Eq.\eqref{37}, i.e.
\be
A_{g(x)}\,A\bigl(\xix\bigr) = A\bigl(\xixp\bigr)\,A\bigl(\Lambda (\xixp,\xix\bigr) .
\label{A6}
\ee

We, finally, remark that the spin boost, $S\bigl(A(\xix)\bigr)$, is given by
\be
S\bigl(A(\xix)\bigr)={1\over{{[2\,R(x)\,\bigl(R(x)-\tilde\xi^5(x)\bigr)]^{\ha}}}} \bigl[R(x)-\tilde\xi^5(x) -\ga^i\ga^5\, \tilde\xi_i(x)\bigl],
\label{A7}
\ee
with the inverse, $S^{-1}\bigl(A(\xix)\bigr)$, being obtained by replacing in \eqref{A7} $\tilde\xi_i(x)$ by $-\tilde\xi_i(x)$.
The Eq.\eqref{A7} can also be written as
\bea
S\bigl(A(\xix)\bigr)
&=&{{R(x)}\over{{[2\,R(x)\,\bigl(R(x)-\tilde\xi^5(x)\bigr)]^{\ha}}}} \bigl[1-\ga^a\ga^5{{\tilde\xi_a(x)}\over{R(x)}}\bigr]
\nonumber \\
 &=&{{R(x)}\over{{[2\,R(x)\,\bigl(R(x)-\tilde\xi^5(x)\bigr)]^{\ha}}}} \bigl[1-\ga^5(\xix)\,\ga^5\bigr],
\label{A8}
\eea
with $\ga^5S\bigl(A(\xix)\bigr)\ga^5{=}-S^{-1}\bigl(A(\xix)\bigr)$ and $S\bigl(A(\xicirc)\bigr){=}{\bf 1}$. 

\newpage

\section{Appendix B}

\qquad The section $\xia$ on $\Et$ is a constrained field in the de Sitter gauge theory as formulated in the embedding
space $R_{4,1}$. The constraining relation \eqref{33}, i.e. $\xia\xib\eta_{ab}+R^2(x)=0$, can be considered
in the variation of the fields $\xia$ in $\L_h$ with the help of a Lagrangean multiplyer, $\bar a$, multiplying a
Weyl invariant addition $\L '$ to $\L_h$ taking the constraint into account. We use for $\L'$ the expression
\be
\L'=\bar a K\sqrt{-g}\,{{\Phi^2(x)}\over{R^4(x)}}\,\Bigl(\xia\,\xib\,\eta_{ab} + R^2(x)\Bigr)
\label{B1}
\ee   
which is Weyl invariant and has the same length dimension as $\L_h$. Physically the addition of \eqref{B1} means
adding a vanishing invariant quantity to the Lagrangean $\L_h$. Varying the multiplyer $\bar a$ one reobtains the 
constraint, as usual; and in the end the constant $\bar a$ is set equal to 1.

In order to derive field equations for the de Sitter fields $\xia$ we now consider the Lagrangean $\L_h{+}\L'$
with $\L_h$ given by Eq.\eqref{348} and go over to the NL form for the de Sitter spinor fields $\pp$ in \eqref{348}
in using Eq.\eqref{331} and shifting the spin boosts $S^{-1}\bigl(A(\xix)\bigr)$ and $S\bigl(A(\xix)\bigr)$
originating from the $\ga(\tilde\xi)$-matrices to the right and left, respectively, yielding, for example, for
$\pp_L(x)$ :
\bea
S^{-1}\bigl(A(\xix\bigr)\,\pp_L(x)&=&\ha(1-i\ga_5)\,S^{-1}\bigl(A(\xix)\bigr)\,\ppx  \\
\nonumber
&=&\ha(1-i\ga_5)\,\pp\xxix\equiv\pp_L\xxix ,
\label{B2}
\eea
and similarly for $\ppb_L(x)$ and the other spinor fields $\pp^u_R(x)$, $\pp^d_R(x)$ and their adjoints.
The resulting form of Eq.\eqref{348} exhibits the $\xix$-dependences explicitly: Only the $\l$-term has a remaining 
dependence on $\xia$ which was already taken into account in the discussion of Sect.\,4. All the other terms
constructed with the help of the nonlinearly transforming or "contact point" fields, $\pp_{L,R}\xxix$, contain no
dynamically relevant $\xix$-dependences despite the (somewhat misleading) notation which was introduced
above (see Eq.\eqref{313}) to denote the NL transformation behaviour under de Sitter gauge transformations.
We thus come to the conclusion that the term denoted by $\delta'\L_h$ in Eq.\eqref{44} is in fact zero. 

\end{appendix}

\vspace{1cm}
\noindent{\bf Acknowledgment}
\vskip3.0mm

I am grateful to Peter Breitenlohner and Heinrich Saller for discussions.

\newpage
%====================================================================
%REFERENCES *******************************************************


\begin{thebibliography}{99}
\bibitem{1} W. Drechsler, Mass Generation by Weyl Symmetry Breaking, \\{\it Found. Phys.} {\bf 29}, 1327-1369,
                   (1999).  [\,Referred to as  I  ]  

\bibitem{2} W. Drechsler and H. Tann, Broken Weyl-Invariance and the Origin of Mass,
                   {\it Found.Phys.} {\bf 29}, 1023- 1064, (1999). 
            
\bibitem{3} W. Drechsler, Gauge Theory for Extended Elementary Objects, \\{\it Class. Quantum Grav.}
                   {\bf 6}, 623-657 (1989).   [\,Referred to as  II  ]

\bibitem{3a}T. Fulton, F. Rohrlich and L. Witten, Conformal Invariance in Physics, {\it Rev. Mod. Phys.}
                    {\bf 34}, 442-457 (1962).

\bibitem{3b} H. Tann,  Einbettung der Quantentheorie eines Skalarfeldes in eine \\ Weyl-Geometrie --
                    Weyl Symmetrie und ihre Brechung.  \\
                   {\it Dissertation der Fakult\"at f\"ur Physik der LMU - M\"unchen},  April 1997.

\bibitem{4} S. Weinberg, A Model of Leptons, {\it Phys. Rev. Lett.} {\bf 19},
                   1264-1266   (1967).

\bibitem{5} S. Weinberg, The Making of the Standard Model, Talk given at CERN on September 16, 2003. 
                   arXiv:hep-ph/0401010v1

\bibitem{6} M. Gell-Mann, Symmetry of Baryons and Mesons, {\it Phys. Rev.} {\bf 125}, 1067-1084    (1962).

\bibitem{7} W. Drechsler, Poincar$\acute{\rm e}$ Gauge Theory, Gravitation, and Transformation of Matter Fields, 
                   {\it Fortschr. Phys.} {\bf 32}, 449-472  (1984).
\bibitem{8} W. Drechsler, Wave Equations on a de Sitter Fiber Bundle, \\{\it Fortschr. Phys.} {\bf 23}, 607-647 (1975).
 
\bibitem{9} W. Drechsler, Soldered Bundles in Particle Physics, \\{\it Fortschr. Phys.} {\bf 38}, 63-75  (1990). 
 
\bibitem{10} W. Drechsler, Group Contraction in a Fiber Bundles with Cartan Connection,
                     {\it J. Math. Phys.} {\bf 18}, 1358-1366  (1977).

\bibitem{11} W. Drechsler, Modified Weyl Theory and Extended Elementary Objects, {\it Found. Phys.} {\bf 19},
                     1479-1497  (1989).

\bibitem{12} W. Drechsler and W. Thacker, Generalized Spinor Fields and Gravitation, {\it Class. Quantum Grav.} {\bf 4},
                     291-318  (1987).                           

\bibitem{13} S. Kobayashi and K. Nomizu, {\it Foundations of Differential
           Geometry}, Vol. I, New York, Interscience  1963.


\end{thebibliography}
\end{document}